\documentclass[twocolumn,numberedappendix,iop]{emulateapj}
\shorttitle{The Collisional Evolution of Debris Disks}
\shortauthors{G\'asp\'ar et al.}
\submitted{\today}
\journalinfo{The Astrophysical Journal}

\usepackage{apjfonts}
\usepackage{amsmath}
\usepackage{multirow}

\usepackage{lscape}

\begin{document}

\title{The Collisional Evolution of Debris Disks}

\author{Andr\'as G\'asp\'ar$^1$}
\author{George H.~Rieke$^1$}
\author{Zolt\'an Balog$^2$}
\affil{1) Steward Observatory, University of Arizona, Tucson, AZ, 85721\\
2) Max-Plank Institut f\"ur Astronomie, K\"onigstuhl 17, D-69117 Heidelberg, Germany\\
agaspar@as.arizona.edu, grieke@as.arizona.edu, balog@mpia.de}

\begin{abstract}
We explore the collisional decay of disk mass and infrared emission in debris disks. 
With models, we show that the rate of the decay varies throughout the evolution of the disks, increasing 
its rate up to a certain point, which is followed by a leveling off to a slower value. The total 
disk mass falls off $\propto t^{-0.35}$ at its fastest point (where $t$ is time) for our reference model, 
while the dust mass and its proxy -- the infrared excess emission -- fades significantly faster 
($\propto t^{-0.8}$). These 
later level off to a decay rate of $M_{\rm tot}(t) \propto t^{-0.08}$ and 
$M_{\rm dust}(t)~{\rm or}~L_{\rm ir}(t) \propto t^{-0.6}$. This is slower than the 
$\propto t^{-1}$ decay given for all three system parameters by traditional analytic models.

We also compile an extensive catalog of {\it Spitzer} and {\it Herschel} 24, 70, and 100 
$\micron$ observations. Assuming a log-normal distribution of initial disk masses, 
we generate model population decay curves for the fraction of stars harboring debris disks detected 
at 24 $\mu$m. We also model the distribution of measured excesses at the 
far-IR wavelengths (70--100 $\micron$) at certain age regimes. We show general agreement 
at 24 $\micron$ between the decay of our numerical collisional population synthesis model 
and observations up to a Gyr. We associate offsets above a Gyr to stochastic events in a few 
select systems. We cannot fit the decay in the far infrared convincingly with grain strength 
properties appropriate for silicates, but those of water ice give fits more consistent with the 
observations (other relatively weak grain materials would presumably also be successful). 
The oldest disks have a higher incidence of large excesses than predicted by the model;
again, a plausible explanation is very late phases of high dynamical activity around a small
number of stars.

Finally, we constrain the variables of our numerical model by comparing the evolutionary trends 
generated from the exploration of the full parameter space to observations. Amongst other results, 
we show that erosive collisions are dominant in setting the timescale of the evolution and that 
planetesimals on the order of \mbox{100 km} in diameter are necessary in the cascades for our 
population synthesis models to reproduce the observations. 

\end{abstract}

\section{Introduction}

Planetary debris disks provide the most accessible means to explore the outer zones of planetary 
systems over their entire age range -- from 10 Gyr to examples just emerging from the formation 
of the star and its planets at 10 Myr. Debris disks are circumstellar rings of dust, rocks, and 
planetesimals, which become visible in scattered light and infrared emission because of their 
large surface areas of dust. Because this dust clears quickly, it must be constantly replenished 
through collisions amongst the larger bodies, initiated by the dynamical perturbing forces of 
nearby planets \citep{wyatt08}. Thus, the presence of a debris disk signals not only that the 
star has a large population of planetesimals, but that there is possibly
at least one larger body to stir this population \citep{kenyon01,mustill09,kennedy10}. The overall 
structures of these systems are indicative of the 
processes expected to influence the structures of the planetary systems. They result from 
sublimation temperatures and ice lines \citep[e.g.,][]{morales11} and sculpting by unseen 
planets \citep[e.g.,][]{liou99,quillen02,kuchner03,moran04,moro-martin05,chiang09}, as well 
as from conditions at the formation of the planetary system. 

However, debris disks undergo significant evolution \citep{rieke05,wyatt08}. Studies of other 
aspects of disk behavior, such as dependence on metallicity or on binarity of the stars, 
generally are based on stars with a large range of ages, and thus the evolution must be taken 
into account to reach reliable conclusions about the effects of these other parameters. Analytic 
models of the collisional processes within disks have given us a rough understanding of their 
evolution \citep{wyatt07,wyatt08}, yielding decays typically inversely with time for the steady state (constant rate of 
decay). Multiple observational programs have characterized the decay of debris disks 
\citep[e.g.,][]{spangler01,greaves03,liu04,rieke05,moor06,siegler07,gaspar09,carpenter09,moor11} 
and indicate general agreement with these models. However, these comparisons are limited by small 
sample statistics, uncertainties in the stellar ages, and the difficulties in making a 
quantitative comparison between the observed incidence of excesses and the model predictions.  

In fact, more complex numerical models of single systems 
\citep{thebault03,lohne08,thebault07,gaspar12a} have shown that the decay is better described 
as a quasi steady state, with rates varying over time rather than the simple decay slope of 1 
typically found in traditional analytic models. \cite{lohne08} present the evolution of debris 
disks around solar-type stars (G2V), using their 
cascade model {\tt ACE}. They yield a dust mass decay slope of 0.3--0.4. 
The models of \cite{kenyon08} yield a fractional infrared 
luminosity ($f_d = L_{\rm dust}/L_{\ast}$) 
decay slope between 0.6 and 0.8. The latest work presented by \cite{wyatt11} indicates an 
acceleration in dust mass decay, with the systems initially losing dust mass following a decay 
slope of 0.34, which steepens to 2.8 when Poynting-Robertson drag becomes dominant. For the same 
reasons as with the analytic models, these predictions are inadequately tested against the 
observations. We summarize the decay slopes determined by observations and models in
\mbox{Table \ref{tab:comp}}.

\setcounter{table}{0}
\begin{deluxetable*}{lllll}
\tabletypesize{\scriptsize}
\tablewidth{0pt}
\tablecolumns{5}
\tablecaption{The decay trends in the literature, with proportionality of variables to time given
 as $\propto t^{-\xi}$\label{tab:comp}}
\tablehead{
\colhead{Paper}					& \colhead{$M_{\rm tot}(t)$}	& 
\colhead{$f_d(t)$ or $f_{d(24)}(t)$ or $M_{\rm dust}(t)$}	& \colhead{Exc (\%)} 	& \colhead{Notes}}
\startdata
\multicolumn{5}{c}{Observations of ensembles of debris disks} \\
\hline\hline\\
\cite{silverstone00}\dotfill			&  			& $\xi = 1.75$       			&	& Average $f_d$ fitted (clusters)\\
\cite{spangler01}\dotfill			&  			& $\xi = 1.76$       			&	& Average $f_d$ fitted (clusters)\\
\multirow{2}{*}{\cite{greaves03}\dotfill}	&  			& \multirow{2}{*}{$\xi \le 0.5$\tablenotemark{$\ast$}} 	&	& Calculated from excess fractions assuming\\
						&  			& 					&	& a constant distribution of dust masses \\
\cite{liu04}\dotfill				&  			& $\xi = 0.7$\tablenotemark{$\ast$}	&	& Upper envelope of submm disk mass decay\\
\cite{rieke05}\dotfill				&  			& $\xi = 1.0$				&	& {\it Spitzer} MIPS [24] fraction\\
\cite{gaspar09}\dotfill				&  			& 					& $\xi = 0.43$	      & Fitted published data between 10 -- 1000 Myr\\
\cite{moor11}\dotfill				&  			& $\xi = 0.3 \dots 1.0$ 			&	& Dispersion between these extremes \\
\hline\\
\multicolumn{5}{c}{Analytic models of single debris disk evolution}\\
\hline\hline\\
\cite{spangler01}\dotfill			& $\xi = 2.0$ 		& $\xi = 2.0$\tablenotemark{$\ast$}	&	& Assumed steady-state\\
\cite{dominik03}\dotfill			& 			& $\xi = 2.0$ 				&	& Collision dominated removal \\
\cite{dominik03}\dotfill			& 			& $\xi = 1.0$ 				&	& PRD dominated removal \\
\cite{wyatt07}\dotfill				& $\xi = 1.0$		& $\xi = 1.0$\tablenotemark{$\ast$}	&	& Assumed steady-state \\
\hline\\
\multicolumn{5}{c}{Numerical models of single debris disk evolution}\\
\hline\hline\\
\cite{thebault03}\dotfill			& $\xi = 0.05$		& $\xi = 0.38$\tablenotemark{$\ast$}	&	& Fitted between 3 and 10 Myr \\
\cite{lohne08}\dotfill				& $\xi = 0.2$		& $\xi = 0.3 \dots 0.4$		&	& \\
\cite{kenyon08}	\dotfill			&			& $\xi = 0.6 \dots 0.8$		&	& \\
\cite{wyatt11}\dotfill  			& $\xi = 0.94$		&					&	& Above 100 Myr \\
\cite{wyatt11}\dotfill 				&			& $\xi = 0.34$\tablenotemark{$\ast$}	&	& Below 200 Myr \\
\cite{wyatt11}\dotfill 				&			& $\xi = 0.97$\tablenotemark{$\ast$}	& 	& Above 2 Gyr \\
\cite{wyatt11}\dotfill 				&			& $\xi = 2.8$\tablenotemark{$\ast$}	&	& PRD dominated above 10 Gyr \\
This work (valid for all systems)\dotfill	& $\xi = 0.33$		& $\xi = 0.8$\tablenotemark{$\ast$}	& 	& At their fastest point in evolution\\
This work (valid for all systems)\dotfill 	& $\xi = 0.08$		& $\xi = 0.6$\tablenotemark{$\ast$}	& 	& At very late ages (quasi steady state)\\
\hline\\
\multicolumn{5}{c}{Population synthesis numerical models of debris disk evolution\tablenotemark{$\dagger$}}\\
\hline\hline\\
This work (early types at 24 $\micron$)		&			&			& $\xi = 0.1$		& 10 -- 250 Myr\\
This work (early types at 24 $\micron$)		&			& 			& $\xi = 2.5$		& 0.4 -- 1 Gyr\\
This work (solar types at 24 $\micron$)		&			&			& $\xi = 0.1$		& 10 -- 100 Myr\\
This work (solar types at 24 $\micron$)		&			& 			& $\xi = 2.6$		& 0.2 -- 0.4 Gyr\\
This work (solar types at 24 $\micron$)		&			& 			& $\xi = 1.4$		& 0.6 -- 10 Gyr
\enddata
\tablenotetext{$\ast$}{Decay timescale calculated for dust mass.}
\tablenotetext{$\dagger$}{Disks placed at radial distances with disk mass distributions as described in Section \ref{sec:fit}. The decay describes the evolution of a disk population and not that of a single disk.}
\end{deluxetable*}

In this paper, we compute the evolution of debris disk signatures in the mid- and far-infrared, 
using our numerical collisional cascade code \citep[][Paper I hereafter]{gaspar12a}. We examine in detail the 
dependence of the results on the model input parameters. We then convert the results into 
predictions for observations of the infrared excesses using a population synthesis routine. 
We compare these predictions with the observations; most of the results at 24 $\micron$ (721  
solar- and 376 early-type stars) are taken from the literature, but in the far infrared we 
have assembled a sample of 430 late-type systems with archival data from Spitzer/MIPS at 70 $\micron$ and 
Herschel/PACS at 70 and/or 100 $\micron$. We have taken great care in estimating the ages of these stars. We find 
plausible model parameters that are consistent with the observations. This agreement 
depends on previously untested aspects of the material in debris disks, such as the 
tensile strengths of the particles. Our basic result confirms that of \cite{wyatt07} 
that the overall pattern of disk evolution is consistent with evolution from a log-normal 
initial distribution of disk masses. It adds the rigor of a detailed numerical 
cascade model and reaches additional specific conclusions about the placement of the 
disks and the properties of their dust.

Although our models generally fit the observed evolution well, there is an excess of debris 
disks at ages greater than 1 Gyr, including systems such as 
HD 69830, $\eta$ Crv, and BD +20 307. We attribute these systems 
to late-phase dynamical shakeups in a small number of planetary systems. In support of this 
hypothesis, a number of these systems have infrared excesses dominated by very small dust 
grains \citep[identified by strong spectral features;][]{beichman05,song05,lisse12}. The dust around 
these stars is almost certainly transient and must be replenished at a very high rate. For 
example, HD 69830 has been found to have three Neptune-mass planets within one AU of the star 
\citep{lovis06}; they are probably stirring its planetesimal system vigorously.

The paper is organized as follows. In section \ref{sec:model}, we present the decay behavior 
of our reference model in three separate parameter spaces. In section \ref{sec:obs}, we 
introduce a set of carefully vetted observations that we will use to verify our model and 
to constrain its parameters, while in section \ref{sec:fit} we establish a population 
synthesis routine and verify our model with the observed decay trends. In section 
\ref{sec:constr}, we constrain the parameters of our collisional cascade model using 
the observations, and in section \ref{sec:concl}, we summarize our findings. We provide 
an extensive analysis of the dependence of the predicted decay pattern on the model 
parameters in the Appendix.

\section{Numerical modeling of single disk decay}
\label{sec:model}

We begin by probing the general behavior of disk decay, using the reference model presented in 
the second paper of our series \citep[][Paper II, hereafter]{gaspar12b}. Models fitted to the full 
set of observations will be discussed in Sections \ref{sec:fit} and \ref{sec:constr}. We refer the 
reader to Papers I and II for the details of the model variables. We define the dust mass as the 
mass of all particles smaller than 1 mm in radius within the debris ring.  
In the Appendix, we analyze the dependence of the decay of a single disk on system variables also 
using the models presented in Paper II, and show the effects that changes in the model variables 
have on the evolution speed of the collisional cascade and/or its scaling in time.

Our reference model (Paper II), is of a 2.5 AU wide (${\Delta R}/R = 0.1$) debris disk situated 
at 25 AU radial distance around an A0 spectral-type star with a total initial mass of 
1 M$_{\Earth}$. The largest body in the system has a radius of 1000 km. The dust mass-distribution 
of the model, once it reaches a quasi steady state, is well approximated by a power-law with a slope 
of 1.88 (3.65 in size space). In the following subsections, we describe the evolution of the decay 
of this model. We analyze the decay of three parameters: the total mass within the system, the dust mass 
within the system, and -- to verify its decay similarity to that of the dust mass -- the fractional 
24 $\micron$ infrared emission ($f_{d(24)} = F_{\rm disk}(24)/F_{\ast}(24)$). 

\subsection{The decay of the total disk mass}
\label{sec:mtot}

The decay of total disk mass is not observable, as a significant portion of it is concentrated in the 
largest body/bodies in the systems, which do not emit effectively. As we show later, the evolution of 
the total mass is not strongly coupled to the evolution of the observable parameters, which is a 
``double edged sword''. Fitting the evolution of the observables can be performed with fewer 
constraints; however, we learn less about the actual decrease of the system mass when using a model 
that is less strict on including realistic physics at the high mass end of the collisional cascade. 
Also, the long-term evolution of the dust will be affected by the evolution of the largest masses in 
a system, meaning that long-term predictions by models not taking this evolution into account 
correctly may be inaccurate. On the other hand, comparison between different collisional models and 
their collisional prescriptions is enabled by this decoupling.

\begin{figure*}
\begin{center}
\includegraphics[angle=0,scale=1.4]{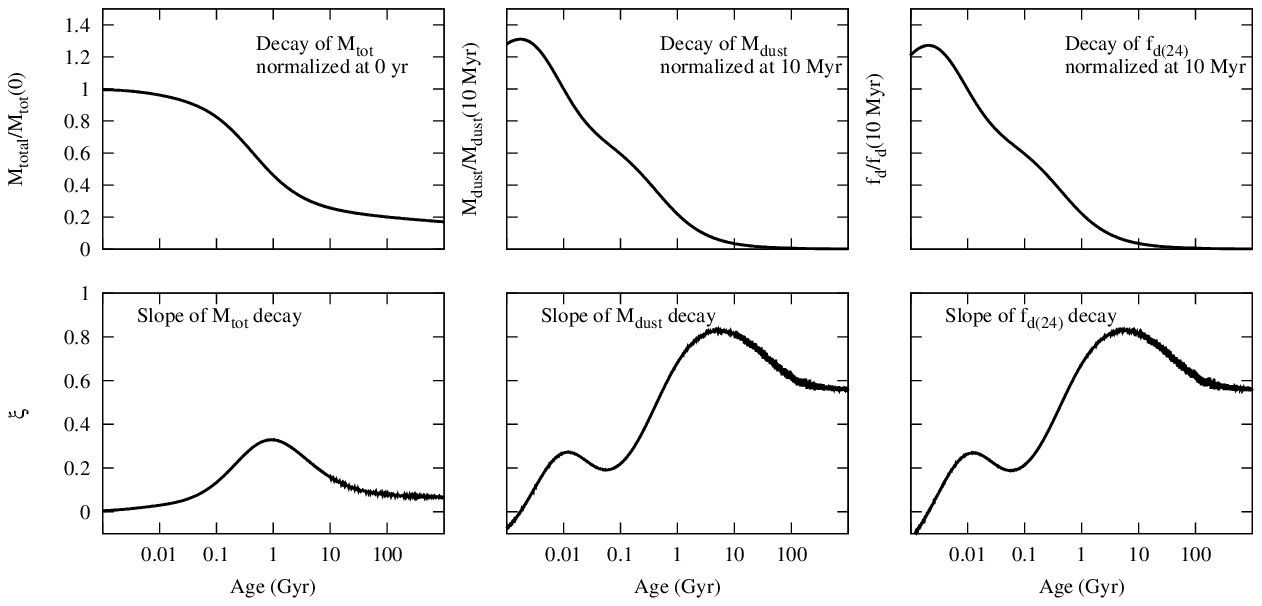}
\caption{The decay of the reference model introduced in Paper II. The top row shows the decay 
of the total mass, the dust mass, and the fractional 24 $\micron$ infrared emission ($f_{d(24)}=F_{\rm disk}(24)/F_{\ast}(24)$); the bottom row shows 
the corresponding decay slopes for each parameter at the same points in time. The plots highlight 
that the decay is not a steady state process.}
\label{fig:Reference}
\end{center}
\end{figure*}

We show the evolution of 
the total disk mass of our reference model in the top 
left and the evolution of the decay slope of the total mass in the bottom left panels of Figure 
\ref{fig:Reference}. The evolution is slow up to 100 Myr (until the larger bodies settle in the quasi 
steady state), after which there is a relatively rapid decay. It reaches its steepest and quickest 
evolution around 1 Gyr, when $\xi \approx 0.35$, where $\xi$ 
is the time exponent of the decay ($\propto t^{-\xi}$). The decay then slows down; 
settling at $\xi \approx 0.08$.
Although Figure 2(c) of \cite{wyatt11} hints at a similar decrease in evolution speeds,
that paper only analyzes the total and dust mass evolution that is proportional to $t^{-0.94}$ and
does not mention a decrease in evolution speeds. Similarly, Figure 8 of \cite{lohne08} possibly hints
at a similar decrease in evolution speeds at the latest stage in evolution, but this behavior is not analyzed in
depth. This discrepancy likely originates from the differing
physics included in the models, such as the omission of erosive collisions and using a continuity equation
for the entire mass range by \cite{lohne08}.

In the Appendix, we show that variations to the total initial disk mass only
scale the decay trend in time (linearly), but not its pattern of evolution, meaning that more massive disks will reach the same
$\xi \approx 0.08$, but at earlier times. Since our reference model is a low-mass disk, the majority of observable 
disks will reach this slow evolutionary state well under a few Gyrs (a disk a hundred times more dense than our 
reference model will settle to its slow decay at $\approx 1~{\rm Gyr}$). This property is used in the population 
synthesis calculations in section \ref{sec:fit}.

\subsection{The decay of the dust mass}

Analytic models of debris disks assume that they are in steady state equilibrium. Under such
assumptions the dust mass decay is proportional to the decay of the total system mass. In reality,
since there is no mass input at the high mass end, the systems evolve in a quasi steady state. 
Since mass evolves downwards to smaller scales within the mass-distribution, the further we move 
away from the high-mass cutoff, the better a steady-state approximation for the collisional cascade 
becomes. This is the reason steady-state approximations for the observed decays have been relatively 
successful, but not exact.

Our model shows a more realistic behavior. We show the evolution of the dust mass in the top middle, 
and the evolution of the 
decay slope of the dust mass in the bottom middle panels of Figure \ref{fig:Reference}. 
Since the final particle mass (size) distribution slope is steeper than the initial
one (Paper II), dust mass will increase in the beginning of the evolution. The evolution
speed increases up to around 0.01 Gyr, after which it stays roughly constant up to 0.1 Gyr. 
This is the period where the larger disk members settle into their respective quasi steady state. 
The evolution once again increases from 0.1 Gyr to a few Gyr, following the formation of the 
``bump'' in the size distribution at larger sizes. The decay slows down again once the entire 
mass range has settled in its quasi steady state, with a decay $\propto t^{-0.6}$.
 
\subsection{The decay of the fractional infrared emission}
\label{sec:lir}

Although our primary interest is the underlying mass and the largest planetesimals in a debris disk, the observable
variable is the infrared emission of the smallest particles. The emission from the debris disk is calculated
following algorithms similar to those in \cite{gaspar12b}. 
For our reference model we assumed a grain composition of astronomical silicates \citep{draine84}, while 
for the icy debris disks introduced in Section \ref{sec:farIR} we assumed a Si/FeS/C/Ice mixture composition 
\citep{min11}. Since our model is currently a 1D particle-in-a-box code, we assumed the modeled size distribution 
to be valid throughout the narrow ring.

We show the evolution of the fractional 24 $\micron$ emission of our reference model in the top right, and the evolution of its 
decay slope in the bottom right panels of Figure \ref{fig:Reference}. We follow the evolution of
the fractional 24 $\micron$ emission instead of the fractional infrared luminosity, as they will be identical in a 
quasi steady state and we avoid integrating the total emission of the disk at each point in time.
The plots clearly show that the evolution of the emission is a proxy for the evolution of the dust mass 
in a system, as its decay properties mirror that of the dust mass. From hereon, we will only focus on the 
evolution of the infrared emission -- which is the observable quantity -- and neglect the dust mass.

\section{Observations}
\label{sec:obs}

We compiled an extensive catalog of 24 -- 100 $\micron$ observations of sources with reliable
photometry and ages from various sources. {\it Spitzer} 24 and 70 $\micron$ data for field
stars were obtained from J.\ Sierchio et al.\ (Submitted to ApJ), \cite{su06}, and
K.\ Y.\ L.\ Su (private communication). We added 24 $\micron$ 
data from a number of stellar cluster studies (see Table \ref{tab:clusters}). Publicly available 
PACS 70 and 100 $\micron$ data from the {\it Herschel} DEBRIS \citep{matthews08,matthews10} and 
DUNES \citep{eiroa10a,eiroa10b,eiroa11} surveys were also obtained from the HSA data archive.
MIPS 24 and 70 $\micron$ data for the stars in these surveys were also added to our analysis. 

\setcounter{table}{2}
\begin{deluxetable*}{llllllll}
\tabletypesize{\footnotesize}
\tablecolumns{8}
\tablewidth{0pt}
\tablecaption{The excess fraction in [24] for early-type stars (A0--A9) and solar-type stars (F5--K9) in clusters/associations.\label{tab:clusters}}
\tablehead{
\colhead{~~~~~~~~~~Name~~~~~~~~~~} & \colhead{Age} & \multicolumn{2}{c}{A0--A9} & \multicolumn{2}{c}{F5--K9} & Excess  & Age \\
& \colhead{[Myr]} & \colhead{[\#]} & \colhead{[\%]} & \colhead{[\#]} & \colhead{[\%]} & \multicolumn{2}{c}{Reference~~~~~~}}
\startdata
$\beta$ Pic MG\dotfill	& 12$^{+8}_{-4}$	& 4/7	& 57.1$^{+14.9}_{-18.0}$	& 3/6	& 50.0$\pm 17.7$ 		& 1	& 2\\
LCC/UCL/US\dotfill	& 10-20			& 42/89	& 47.2$^{+5.3}_{-5.1}$		& 42/92 & 45.7$^{+5.3}_{-5.0}$		& 3	& 4,5,6\\
NGC 2547\dotfill	& 30$\pm 5$		& 8/18	& 44.4$^{+11.7}_{-10.4}$	& 8/20  & 40.9$\pm 10.5$ 		& 7	& 7\\
Tuc-Hor\dotfill		& 30$\pm 5$		& 2/5	& 40.0$^{+21.5}_{-15.6}$	& 0/1   & 0.0$^{+60.0}_{+8.40}$         & 1	& 8\\
IC 2391\dotfill		& 50$\pm 5$		& 3/8	& 37.5$^{+17.9}_{-12.8}$	& 3/10	& 30.0$^{+16.8}_{-10.0}$ 	& 9     & 10\\
NGC2451B\dotfill	& 50$\pm 5$   		& 0/3   & 0.0$^{+36.9}_{+4.2}$          & 6/16  & 37.5$^{+12.9}_{-10.1}$        & 11    & 12\\
NGC2451A\dotfill	& 65$\pm 15$		& 1/5   & 20.0$^{+25.4}_{-7.9}$         & 5/15  & 33.3$^{+13.5}_{-9.5}$         & 11    & 12\\
$\alpha$ Per\dotfill	& 85$^{+5}_{-35}$	& -	& -				& 2/13  & 15.4$^{+14.7}_{-5.3}$  	& 13	& 14,15,16\\
Pleiades\dotfill	& 115$\pm 10$		& 5/26	& 19.2$^{+9.9}_{-5.3}$		& 24/71 & 33.8$^{+6.0}_{-5.0}$	        & 17	& 15,18,19\\
Hyades/Praesepe/Coma Ber \dotfill & 600-800	& 5/46  & 10.9$^{+6.3}_{-3.0}$          & 1/47  &  2.1$^{+4.6}_{-0.6}$	        & 20	& 21,22
\enddata
\tablerefs{
(1) \cite{rebull08};
(2) \cite{ortega02};
(3) \cite{chen11}; 
(4) \cite{preibisch02};
(5) \cite{fuchs06};
(6) \cite{mamajek02};
(7) \cite{gorlova07};
(8) \cite{rebull08}, with arbitrary errors adopted from similar age clusters;
(9) \cite{siegler07}; 
(10) \cite{barrado04};
(11) \cite{balog09};
(12) \cite{hunsch03};
(13) \cite{carpenter09};
(14) \cite{song01}; 
(15) \cite{martin01}
(16) \cite{mamajek08};
(17) \cite{sierchio10};  
(18) \cite{meynet93};
(19) \cite{stauffer98}; 
(20) \cite{urban12};
(21) \cite{gaspar09};
(22) \cite{perryman98}
}
\end{deluxetable*}

\subsection{MIPS 24 $\micron$ data}

At 24 $\micron$, we determined excesses in the MIPS data for field stars by applying an empirical relation between 
${\rm V}-{\rm K}$ and ${\rm K}-[24]$ \citep[see, e.g.,][]{urban12}. We used 2MASS data for the near infrared magnitudes 
for many stars, but where these data are saturated we transformed heritage photometry to the 2MASS 
system \citep[e.g.,][]{carpenter01}. In one case, we derived a K magnitude from COBE data, and in another 
we were forced to use the standard ${\rm V}-{\rm K}$ color for the star, given its spectral type and ${\rm B}-{\rm V}$ color
(both of these cases are identified in Table \ref{tab:midIR}).  We also determined an independent set of estimates of 22 
$\micron$ excesses from the {\it WISE} W3-W4 color. We found that on average this color is slightly offset 
from zero for stars of the spectral types in our study, so we applied a uniform correction of -0.03. 
It is also important that the MIPS 24 $\micron$ and {\it WISE} W4 spectral bands are very similar, with a 
cuton filter at 20 and 19 $\micron$, respectively, and the cutoff determined by the detector response 
(and with identical detector types).  Not surprisingly, then, we found the two estimates of 22 to 24 $\micron$ 
excess to be very similar in most cases; where there were discrepancies, we investigated the photometry 
and rejected bad measurements. We then averaged the two determinations for all stars with measurements 
in both sets. We quote these averages, or the result of a single measurement if that is all that is 
available, in Table \ref{tab:midIR}. Excesses where only {\it WISE} W4 data was available are considered,
but the MIPS 24 $\micron$ field is left empty.

\subsection{MIPS 70 $\micron$ data \label{sec:MIPS70}}

\begin{figure*}
\begin{center}
\includegraphics[angle=0,scale=0.7]{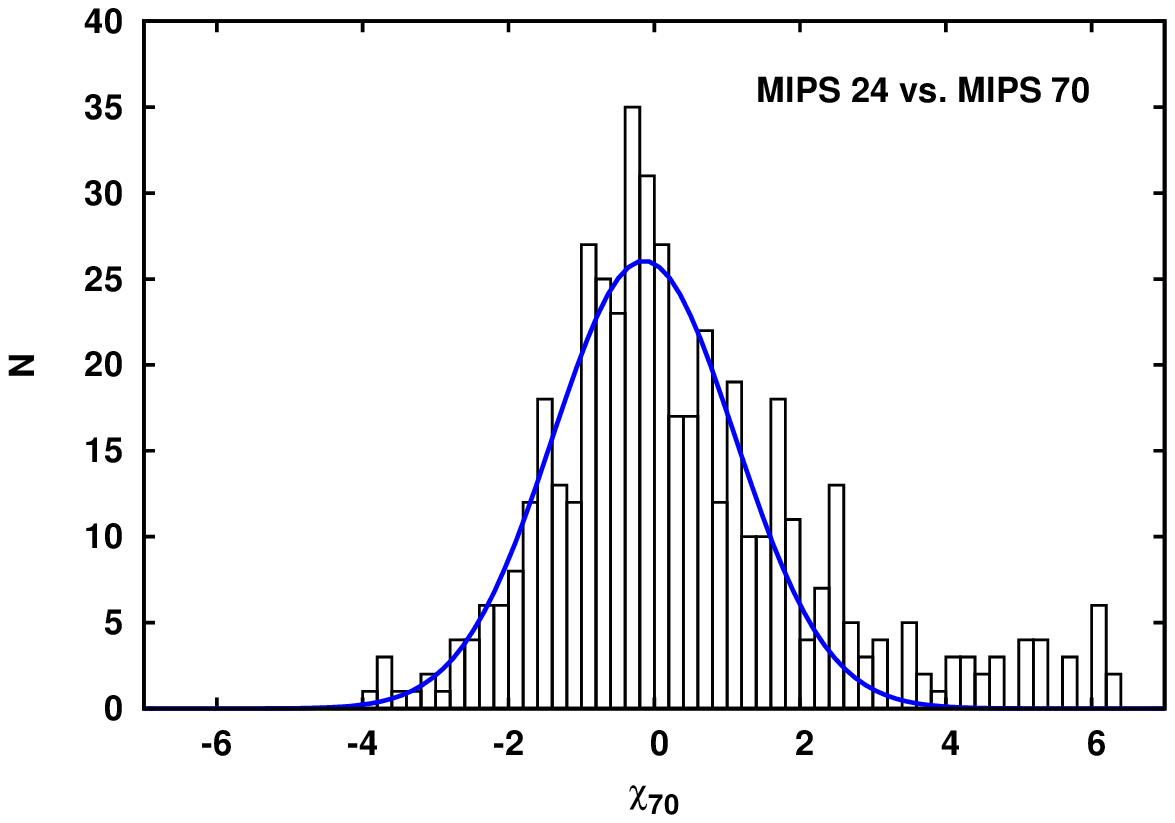}
\includegraphics[angle=0,scale=0.7]{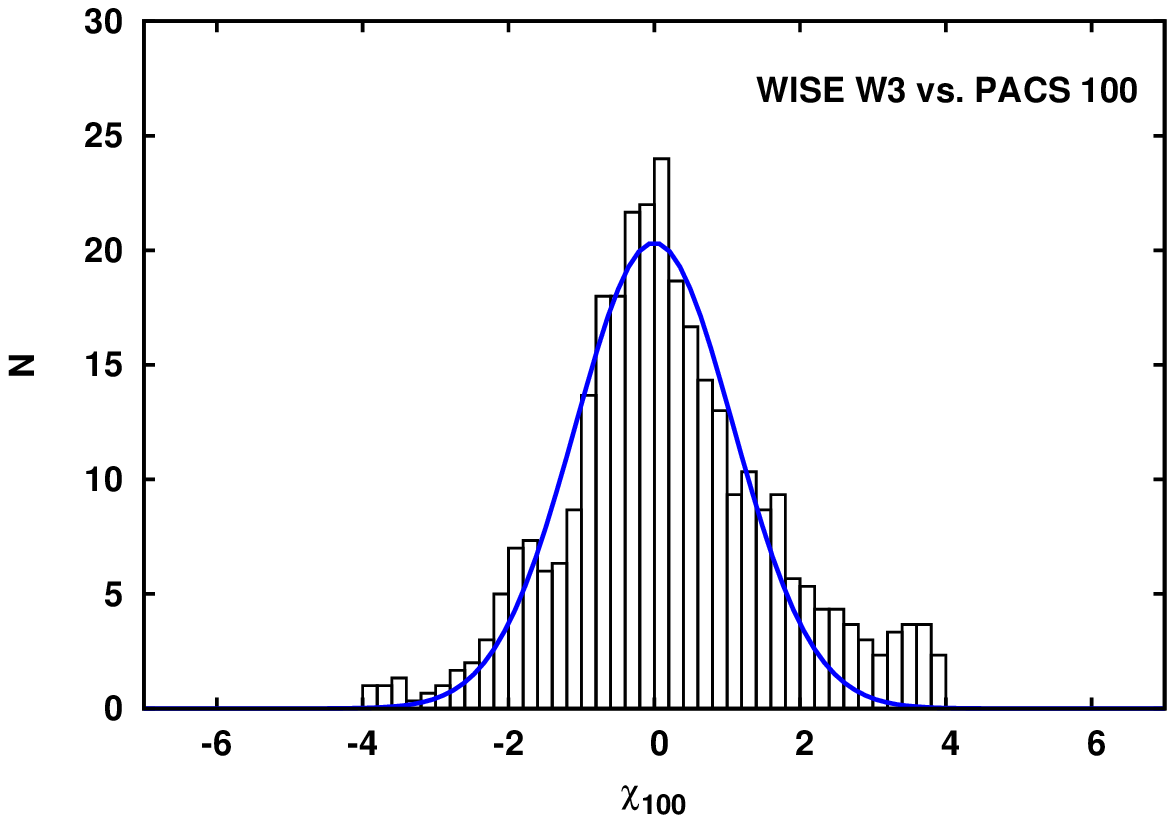}
\caption{Determining the calibration between MIPS 24 and 70 $\micron$, and WISE W3 and PACS 100 $\micron$. 
For displaying the data, the bins were smoothed using a three-bin running average.}
\label{fig:zp}
\end{center}
\end{figure*}

We measured excesses at 70 $\micron$ (MIPS) relative to measurements at 24 $\micron$ (MIPS).
We computed the distribution of the ratio of 24 to 70 $\micron$ flux density, in units of the standard 
deviation of the 70 $\micron$ measurement (we rejected stars with 24 $\micron$ excesses in this 
distribution). The distribution of the ratios of the observed 24 $\micron$ flux density to that at 
70 $\micron$ shows a peak. Because of the range of signal to noise for the stars in the sample, 
this peak is better defined if the ratios are expressed in units of 
standard deviations, or equivalently in terms of the $\chi_{70}$ parameter 
\citep[see, e.g.,][etc.]{bryden06},
\begin{equation} 
\chi_{70} = \frac{F_{70} - P_{70}}{\sigma_{70}}\;,
\end{equation}
where $F_{70}$ is the measured flux density, $P_{70}$ is the predicted flux 
density for the photosphere, and $\sigma_{70}$ is the estimated measurement error. 
The value of $P_{70}$ can be taken to be proportional to the MIPS 24 $\micron$ 
flux density; the proportionality factor of which was adjusted until the peak of the 
distribution was centered around zero. The result, in the left panel of Figure \ref{fig:zp}, 
shows a well defined peak at the photospheric ratio. We fitted the peak with a Gaussian between 
-4 and +2 standard deviations (we did not optimize the fit using larger positive deviations to 
avoid having it being influenced by stars with excesses). This procedure automatically calibrates the 
photospheric behavior, correcting for any overall departure from models, correcting any offsets in 
calibration, and compensating for bandpass effects in the photometry. We used these values to estimate 
the photospheric fluxes at 70 $\micron$. We also corrected the values for excesses at 
24 $\micron$ by multiplying by the excess ratio at this wavelength in all cases where it was 1.10 or 
larger. Smaller values are consistent with random errors and no correction was applied. To test these 
results, we also fitted stellar photospheric models \citep{castelli03} to the full set of photometry 
available for each star from U through MIPS 24 $\micron$ and inspected the behavior of the MIPS 70 
$\micron$ relative to the photospheric levels predicted by these fits. This check neither called into 
question any of the excesses found previously, nor did it suggest additional stars with excesses.

\subsection{PACS 100 $\micron$ data \label{sec:PACS100}}

The {\it Herschel}/PACS data were reduced using the Herschel Interactive Processing 
Environment \citep[HIPE, V9.0 user release,][]{ott10} and followed the 
recommended procedures. We generated the calibrated Level 
1 product by applying the standard processing steps (flagging of bad pixels, flagging of 
saturated pixels, conversion of digital units to Volts, adding of pointing and time 
information, response calibration, flat fielding) and performed second-level deglitching 
with the ``timeordered'' option and a 20 sigma threshold (slightly more conservative 
than the recommended 30 sigma) to remove glitches. This technique 
uses sigma-clipping of the outlying flux values on each map pixel and is very effective for 
data with high coverage. After this stage the science frames were selected from the timeline 
by applying spacecraft-speed selection criteria (as recommended in the pipeline 
script, 18\arcsec/s $<$ speed $<$ 22 \arcsec/s). 
The 1/f noise was removed using high-pass filtering with a filter width of 20 for the 100 
$\mu$m data. This method is based on highpass median window subtraction; thus the 
images might suffer from loss of flux after applying the filter. To avoid this we used a mask with 
20\arcsec~ radius at the position of our sources. After high-pass filtering we combined the 
frames belonging to the two different scan directions and generated the final Level 2 maps 
using photproject also in HIPE. Aperture photometry was performed on the sources using a 
12\arcsec~ radius, while the sky background was determined with an aperture between 
20\arcsec~ and 30\arcsec. Six sub-sky apertures were placed within the nominal sky aperture 
with radii of 12\arcsec, to estimate the variations in the sky background. 
Each image was then inspected. In a few cases, interference by neighboring sources 
caused us to reject the photometry completely; in many more, there was a source in one of 
the six sub-sky apertures and the photometry was checked in place to circumvent the possible influence 
of this source on the results.
Our self-calibration of the data to determine the photospheric level (detailed below) 
circumvents any residual calibration offsets. A summary of the photometry from the DUNES 
and the DEBRIS surveys as well as ages is presented in Table \ref{tab:midIR}.

There is a range of possible choices for the reduction parameters; ultimately, 
the validity of our reduction depends on testing it to see if: 1.) it provides accurate 
calibration; 2.) the noise is well-behaved; 3.) it can be validated against independent 
measurements; and 4.) it is free of systematic errors. We discuss each of these issues 
in turn.
 
In the case of the PACS 100 $\micron$ data, we determined the stellar photospheric ratio of 
WISE W4 22 $\micron$ flux density to that at 100 $\micron$ empirically, following the same
routines as we performed for the MIPS 70 $\micron$ data calibration \citep{gordon07}. We judged the position of the 
peak in the $\chi_{100}$ distribution by Gaussian fitting and found it to be 3\% above a simple 
Rayleigh-Jeans extrapolation.  The far infrared spectral energy 
distributions of stars are not well understood observationally, but theoretical 
models indicate values of 1 - 2\% above Rayleigh Jeans (Castelli, F.\footnote{ 
http://wwwuser.oat.ts.astro.it/castelli/}). For comparison, the absolute calibration at 
MIPS 24 $\micron$ has an uncertainty of 2\% and that of PACS of 
3 - 5\%, so our reduction preserves the calibration to well within its uncertainties.
 
The uncertainties we derive are typical for PACS observations of similar integration 
time. However, a more stringent test is whether they are normally distributed. The 
distribution of $\chi_{100}$ is the distribution of differences from the photospheric flux 
density in units of the estimated standard deviation. As shown in Figure \ref{fig:zp}, it is 
acccurately Gaussian and falls to low levels at the 3-sigma point (the excess above 
3-sigma toward the high end is due to debris-disk infrared excesses. Thus our reduction 
correctly estimates the noise and produces the expected noise distribution.
 
Examination of Table \ref{tab:midIR} shows that the MIPS and PACS measurements are 
generally consistent, as we will demonstrate in more detail below when we discuss identifying 
the members of this sample with detected excess emission. A short summary is that, of 60 
stars with the most convincing evidence for excesses, 56 were observed with both 
telescopes, and for 55 of these there is an indicated signal from each independently 
(> 3 sigma in one and at least 1.4 sigma with the other).
 
Finally, we have tested whether our measurements are subject to systematic errors 
due to missing some extended flux. We set the filtering and aperture photometry 
parameters at values to help capture the flux from extended debris disks. For the 
largest systems known, we still come up 20\% \citep[61 Vir -][]{wyatt12} to 30\% 
\citep[HD 207129 -][]{lohne12} short and we have substituted the values from the 
references mentioned for those we measured. However, for all 15 resolved systems in 
our sample and with studies in the literature \citep{booth13,broek13,matthews10,
liseau10,eiroa10b,kennedy12}, the average underestimate is 6.4\%, and if we exclude 61 Vir and HD 207129 
it is only 3.4\%. This test is severe, since the literature will preferentially contain 
the most dramatic examples of extended disks; in fact, inspecting the DUNES/DEBRIS images 
there are only 2-3 clearly extended systems that are not yet the subject of publications (we note these in Table 
\ref{tab:midIR}). Nontheless, there appears to be little lost flux in our photometry.

\subsection{Determining ages for the field sample stars}

Ages were estimated for these stars using a variety of indicators. 
Chromospheric activity, X-ray luminosity, and gyrochronology as measures of 
stellar age are discussed by \cite{mamajek08}; we used their calibrations. 
To confirm the age estimates past 4 Gyr, we used a metallicity-corrected 
M$_{\rm K}$ vs.\ ${\rm V}-{\rm K}$ 
HR diagram and found excellent correspondence between the assigned ages and the isochrone age. This work is discussed in detail in J.\ Sierchio et al.\ (Submitted to ApJ). 
We also used values of $v \sin i > 10~{\rm km}~{\rm s}^{-1}$ as indicators of youth, and $\log g < 4$
as an indicator of post-main-sequence status (when other indications of youth were absent). 
Our assigned ages and the sources of data that support them are listed in Table \ref{tab:midIR}. We were 
not able to develop a rigid hierarchy among the methods in assigning ages, since occasionally 
an otherwise reliable indicator gives an answer that is clearly not reasonable for a given 
star -- e.g., a low level of chromospheric activity can be indicated for a star whose 
position on the HR diagram is only compatible with a young age; HD 33564 is an example.

\subsection{The decay of planetary debris disk excesses at 24 $\micron$}

{\it Spitzer} 24 $\micron$ data have been used in many studies
of warm debris disk emission \citep[e.g.,][]{rieke05,su06,siegler07,trilling08,gaspar09}.
Given the uncertainties in the ages of field stars, stellar cluster studies, where numerous coeval 
systems can be observed, are strongly favored in disk evolution studies. 
The clusters included in our current research (Table \ref{tab:clusters}) have well defined ages and, more
importantly, homogenic and reliable photometry. Unfortunately, getting an even coverage of ages using 
only clusters is not possible, especially for ages above a Gyr, which is why we combined the stellar 
cluster studies with field star samples. We include the study of 24 $\micron$ excesses
around early-type field stars by \cite{su06}, while the
solar-type stars are included from Sierchio et al.\ (Submitted to ApJ).
We also include the {\it Spitzer} 24 $\micron$ measurements of the sources found in the DUNES and 
DEBRIS {\it Herschel} surveys (K.\ Y.\ L.\ Su, private communication). Our final combined samples have 
721 and 376 sources in the solar-type (F5-K9) and early-type (A0-F5) groups, respectively. 
We summarize our detection statistics in Table \ref{tab:stat}.

\setcounter{table}{3}
\begin{center}
\begin{deluxetable*}{c|c||c|c||c|c||c|c|||c|c}
\tabletypesize{\footnotesize}
\def\arraystrech{6}
\tablecolumns{10}
\tablewidth{0pt}
\tablecaption{The detection statistics of the observational sample. The columns give the detected number of
debris disks over the total number of sources, as a function of age and observed wavelength, for each survey.
The detection criteria are described in the text. \label{tab:stat}}
\tablehead{
\colhead{} & \colhead{Age} & \multicolumn{2}{c}{DUNES}         & \multicolumn{2}{c}{DEBRIS}       & \multicolumn{2}{c}{Additional$^{\dagger}$}    & \multicolumn{2}{c}{Total}\\
\colhead{} & \colhead{(Myr)} & \colhead{24$\micron$} & \colhead{85$\micron^{\ddagger}$} & \colhead{24$\micron$} & \colhead{85$\micron^{\ddagger}$} & \colhead{24$\micron$} & \colhead{85$\micron^{\ddagger}$} & \colhead{24$\micron$} & \colhead{85$\micron^{\ddagger}$}
}
\startdata
\multirow{4}{*}{Early(A0-F5)}  & 1 - 31      & 0/0   & 0/0    & 1/3   & 1/3     & 64/130  & -/-    & 65/133   & 1/3    \\
			       & 31 - 100    & 0/0   & 0/0    & 0/5   & 0/5     & 7/21    & -/-    & 7/26     & 0/5    \\
			       & 100 - 316   & 0/1   & 0/1    & 10/18 & 10/18   & 14/57   & -/-    & 24/76    & 10/19  \\
			       & 316 - 1000  & 0/0   & 0/0    & 7/54  & 8/54    & 9/67    & -/-    & 16/121   & 8/54   \\
			       & > 1000      & 1/3   & 2/3    & 1/17  & 4/17    & -/-     & -/-    & 2/20     & 6/20   \\
\hline
Early Total		       &             & 1/4   & 2/4    & 19/97 & 23/97   & 94/275  & -/-    & 114/376  & 25/101 \\
\hline\hline
\multirow{6}{*}{Solar(F5-K9)}  & 1 - 31      & 0/1   & 0/1    & 0/2   & 0/2     & 58/125  & 2/6    & 58/128   & 2/9    \\
			       & 31 - 100    & 0/0   & 0/0    & 0/1   & 0/1     & 18/57   & 2/3    & 18/58    & 2/4    \\
			       & 100 - 316   & 1/3   & 3/3    & 0/5   & 0/5     & 34/98   & 8/27   & 35/106   & 11/35  \\
			       & 316 - 1000  & 1/16  & 6/16   & 0/30  & 1/30    & 5/86    & 8/39   & 6/132    & 15/85  \\
			       & 1000 - 3160 & 1/33  & 6/33   & 0/34  & 1/34    & 1/32    & 9/32   & 2/99     & 16/99  \\
			       & > 3160      & 1/62  & 10/62  & 0/59  & 5/59    & 0/77    & 5/77   & 1/198    & 20/198 \\
\hline
Solar Total		       &             & 4/115 & 25/115 & 0/131 & 7/131   & 116/475 & 34/184 & 120/721  & 66/430 \\
\hline\hline\hline
Total (A0-K9)		       &             & 5/119 & 27/119 & 19/228 & 30/228 & 210/750 & 34/184 & 234/1097 & 91/531  
\enddata
\tablenotetext{$\dagger$}{Additional data from: J.\ Sierchio et al.\ (Submitted to ApJ), \cite{su06}, Su K.\ Y.\ L.\ (priv. comm.), and
cluster data from Table \ref{tab:clusters}.}
\tablenotetext{$\ddagger$}{The flux at the dummy 85$\micron$ band is calculated as described in Section \ref{sec:farIR}.}
\end{deluxetable*}
\end{center}

\begin{figure}
\begin{center}
\includegraphics[angle=0,scale=0.51]{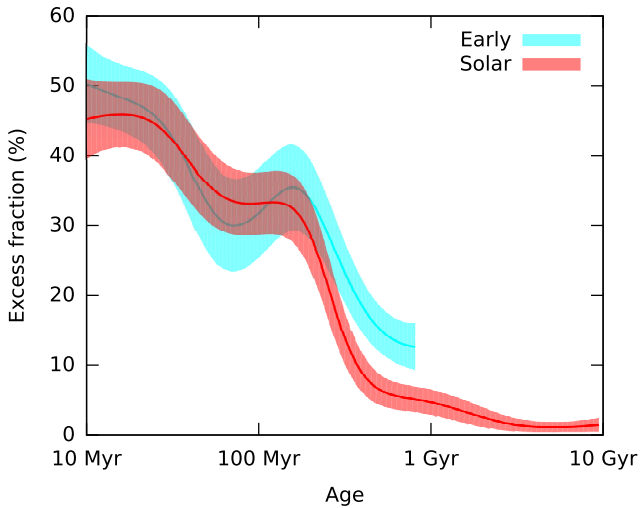}
\caption{The smoothed excess fraction decay curves at 24 $\micron$ for early- and solar-type stars, with
1 $\sigma$ error bars.
The solar-type stars show a slightly quicker decay than the early-types.}
\label{fig:smoothed}
\end{center}
\end{figure}

For our current study, we are interested in the fraction of sources with excess as a function of 
stellar age. We defined a significant excess to occur when the excess ratio (defined as the ratio 
of the measured flux density to the flux density expected from the stellar photosphere) was $> 1.1$
\citep[see, e.g.,][for details of how this threshold is determined]{urban12}.
Classically, sources are binned into age bins and then 
the fraction of sources with excess is determined for each age bin. Instead, we ran a Gaussian 
smoothing function over the observed age range, with a 
Gaussian smoothing width of 0.2 dex in $\log({\rm age})$. With this method, we generate smooth excess 
fraction (defined as the fraction of the sample of stars with excess ratios above some threshold, in 
this case above 1.1) decay curves. Errors of these decay curves were calculated using the method 
described in \cite{gaspar09}. Our final smoothed decay curves at 24 $\micron$ with $\pm 1~\sigma$ errors
for the early- and solar-type stars are shown in Figure \ref{fig:smoothed}. The solar-type stars show a 
slightly quicker decay between 0.1 and 1 Gyr, outside of the 1 $\sigma$ errors. We compare 
these decay curves with population synthesis models in Section \ref{sec:fit}.

\subsection{The decay of planetary debris disk excesses at 70--100 $\micron$}

The MIPS 70 and PACS 100 $\micron$ data are suitable for following the evolution of cold debris
disks \citep{rieke05,su06,wyatt08}. The observations are inhomogenous, having non-uniform detection limits,
which are frequently significantly above the stellar photospheric values. Due to
this, unfortunately, a coherent disk fraction decay can not be calculated, such as for the 24 $\micron$ observations.
We have developed new methods on analyzing the decay of the cold debris disk population, which we detail in 
Section \ref{sec:farIR}.

We used the combined MIPS/PACS far infrared data to generate a reliable list of stars with far infrared 
excess emission. First, there are 35 stars with {\it both} $\chi_{70}$ and $\chi_{100} > 4$ and 4 
more measured only with PACS with $\chi_{100} > 4$ (3 of them have $\chi_{100} > 10$). Thirteen
additional stars have $\chi$ measured with one telescope $\ge 4$ {\it and} $\chi$ with the other telescope $> 2$. These 
52 stars should constitute a very reliable ensemble of far infrared excesses. The remaining eight candidates are 
HD 7570 ($\chi_{70} = 3.9$; $\chi_{100} = 2.6$);  HD 23281 ($\chi_{70} = 8.6$ and $\chi_{100} = 1.8$), HD 87696 
($\chi_{70} = 3.0$ and $\chi_{100} = 2.4$), HD 88955 ($\chi_{70} = 4.0$ and $\chi_{100} = 1.6$), HD 223352 
($\chi_{70} = 6.0$ and $\chi_{100} = 0.2$), HIP 72848 ($\chi_{70} = 2.3$ and $\chi_{100} = 3.2$). 
HIP 98959 ($\chi_{70} = 3.7$ and $\chi_{100} = 2.4$), and HIP 107350 ($\chi_{70} = 4.5$ and $\chi_{100} = 1.4$). 
In all these cases, there is a strong case for a detected excess with a promising indication of a far infrared excess 
with each telescope (excepting for HD 223352), so we add them to the list of probable excesses for a total of 60. 
Finally, HD 22001 has $\chi_{70} = 0.85$, $\chi_{100} = 8.3$; inspection of the measurements indicates that the 
probably far infrared spectrum does indeed rise steeply from 70 to 100 $\micron$. This behavior is expected of a 
background galaxy; in general the spectral energy distributions of debris disks fall (in frequency units) from 
70 to 100 $\micron$. We therefore do {\it not} include this star in our list of those with probable debris disk 
excesses. Excluding this last star, the total combined DEBRIS/DUNES sample has 373 members. Of these, 347 are within
our modeled spectral range (A0-K9) and have age estimates, of which 57 have probable debris disk excesses.

By comparing the results from both MIPS and PACS and also maintaining 
their independence, we have been able to identify reliably a set of stars with far infrared excess emission. 
We list the final photometric data for these sources in Table \ref{tab:midIR}. For our current study we also 
include the 70 $\micron$ measurements of Sierchio et al.\ (Submitted to ApJ). Our final 
catalog of far-IR measurements totals 557 sources, of which 531 are within our modeled
spectral range (A0-K9) and have age estimates (101 early and 430 solar-type). However, we do not analyze 
the decay of far IR excess emission around early-type stars due to the intrinsic lack of data past 1 Gyr.
The observational statistics on the far-IR sample can also be found in Table \ref{tab:stat}.

\section{Population synthesis and comparison to observations}
\label{sec:fit}

In this section, we compare the decay of infrared excesses predicted by our model, using
population synthesis, to the observed fraction of sources with excesses at $24~\micron$ 
and to the distribution of excesses at 70 -- 100 $\micron$. The two wavelength regimes are
dealt with differently due to reasons explained in Section \ref{sec:obs}.

\subsection{Disk locations}
\label{sec:location}

By fitting black body emission curves to IRS spectral energy distributions (SEDs), 
\cite{morales11} found that the majority of debris disks have just a cold component or separate cold 
and warm components. Mostly independent of stellar spectral type, the respective blackbody temperatures for the 
warm and cold components yielded similar values.

\begin{figure*}
\begin{center}
\includegraphics[angle=0,scale=0.54]{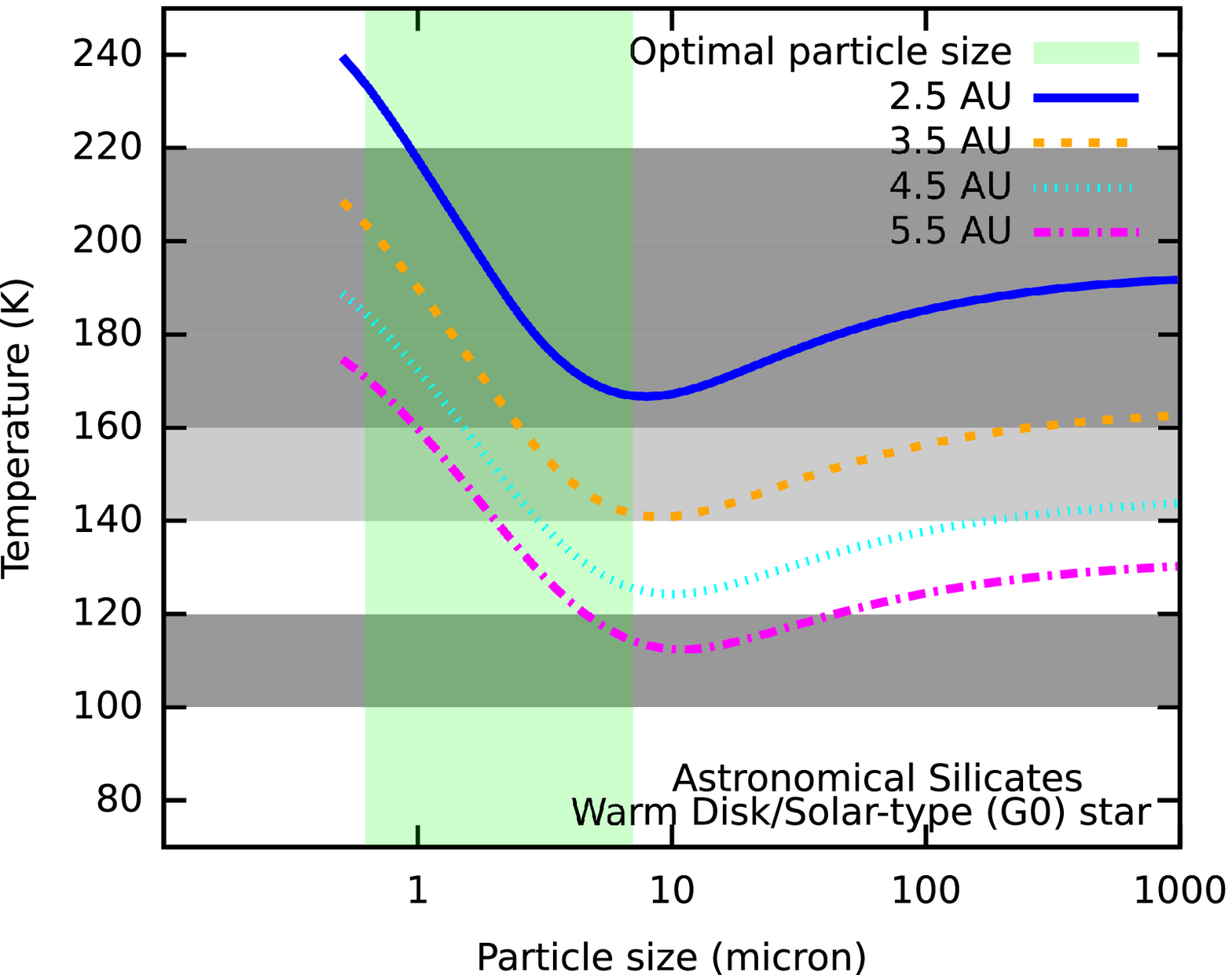}
\includegraphics[angle=0,scale=0.54]{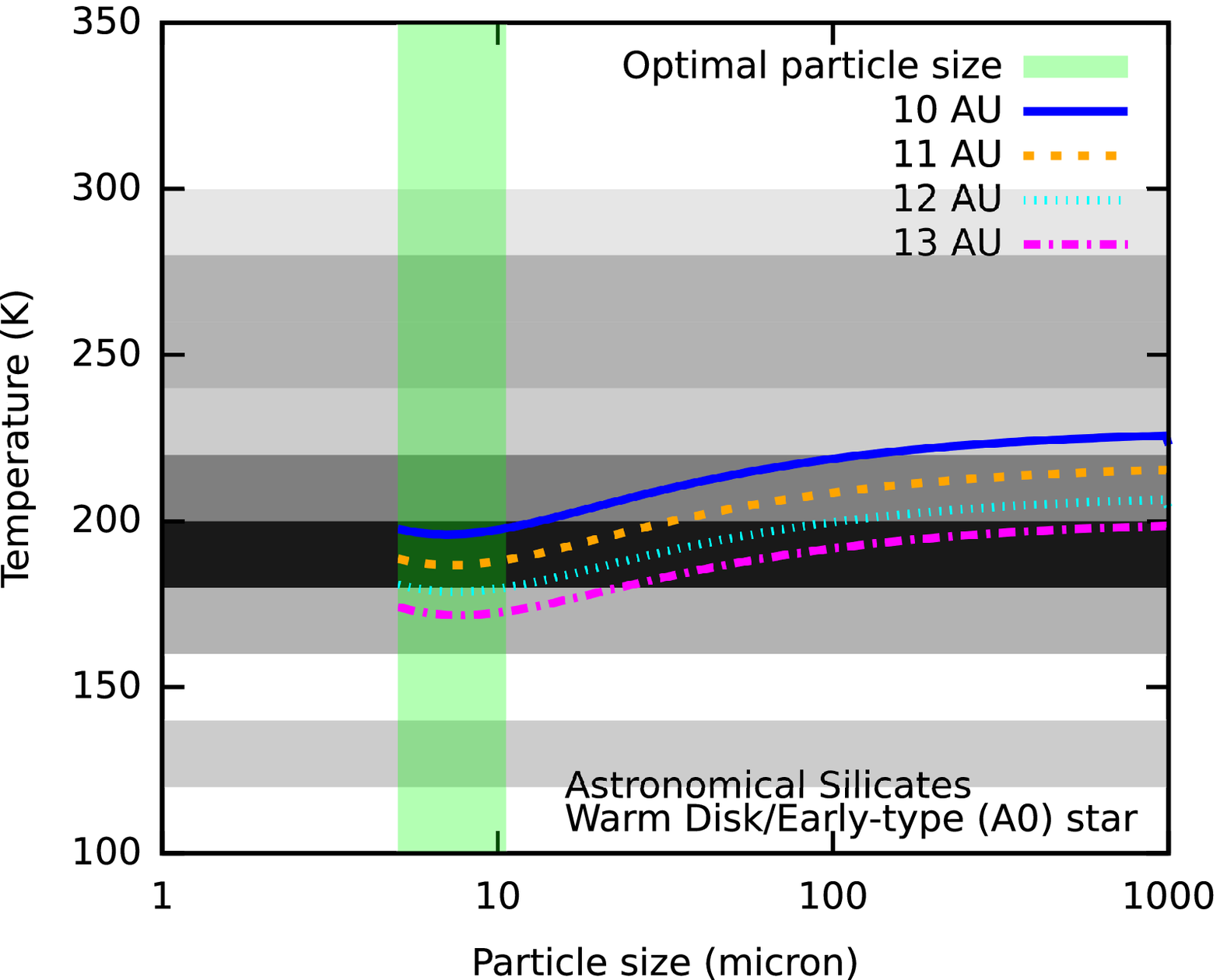}

\includegraphics[angle=0,scale=0.54]{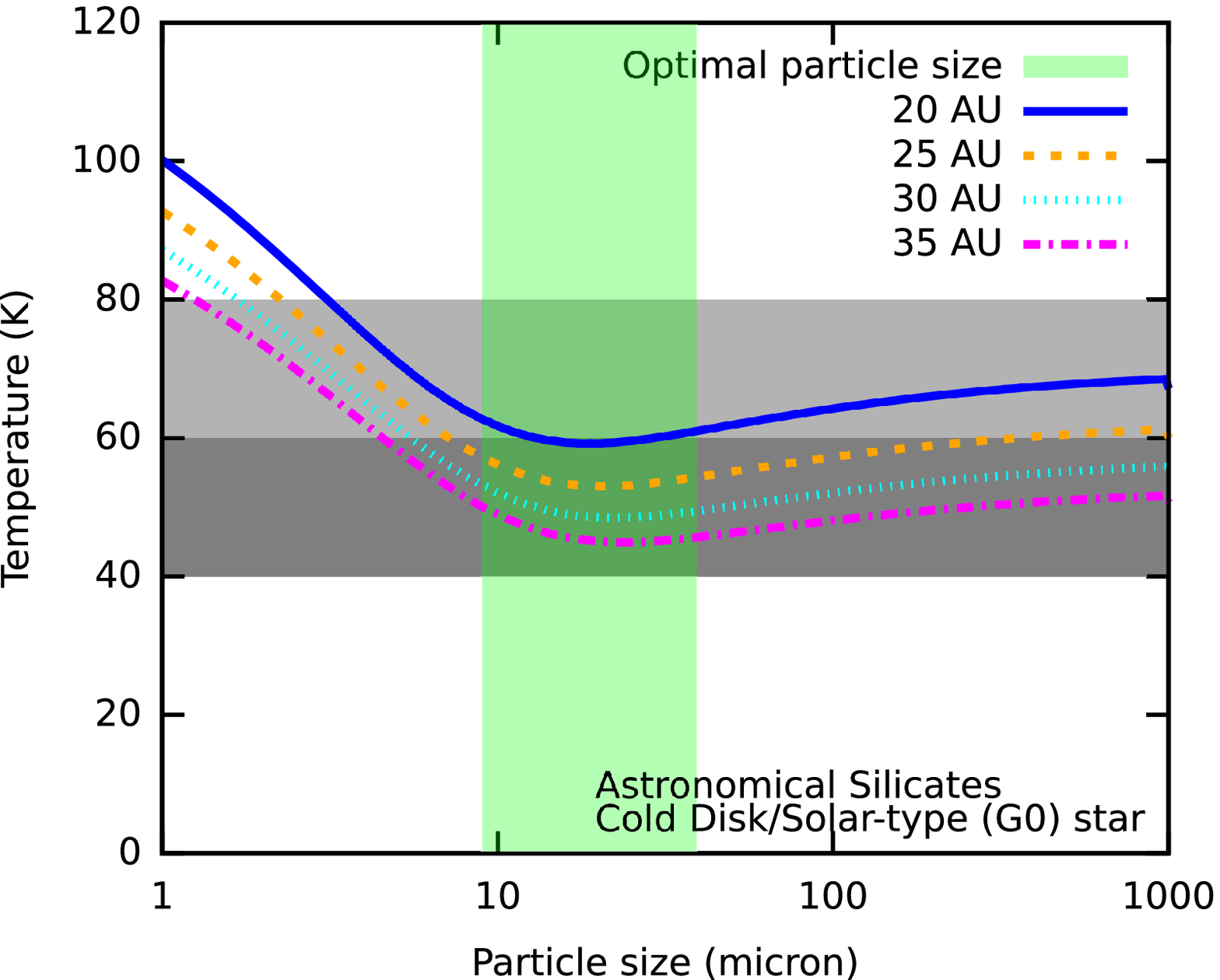}
\includegraphics[angle=0,scale=0.54]{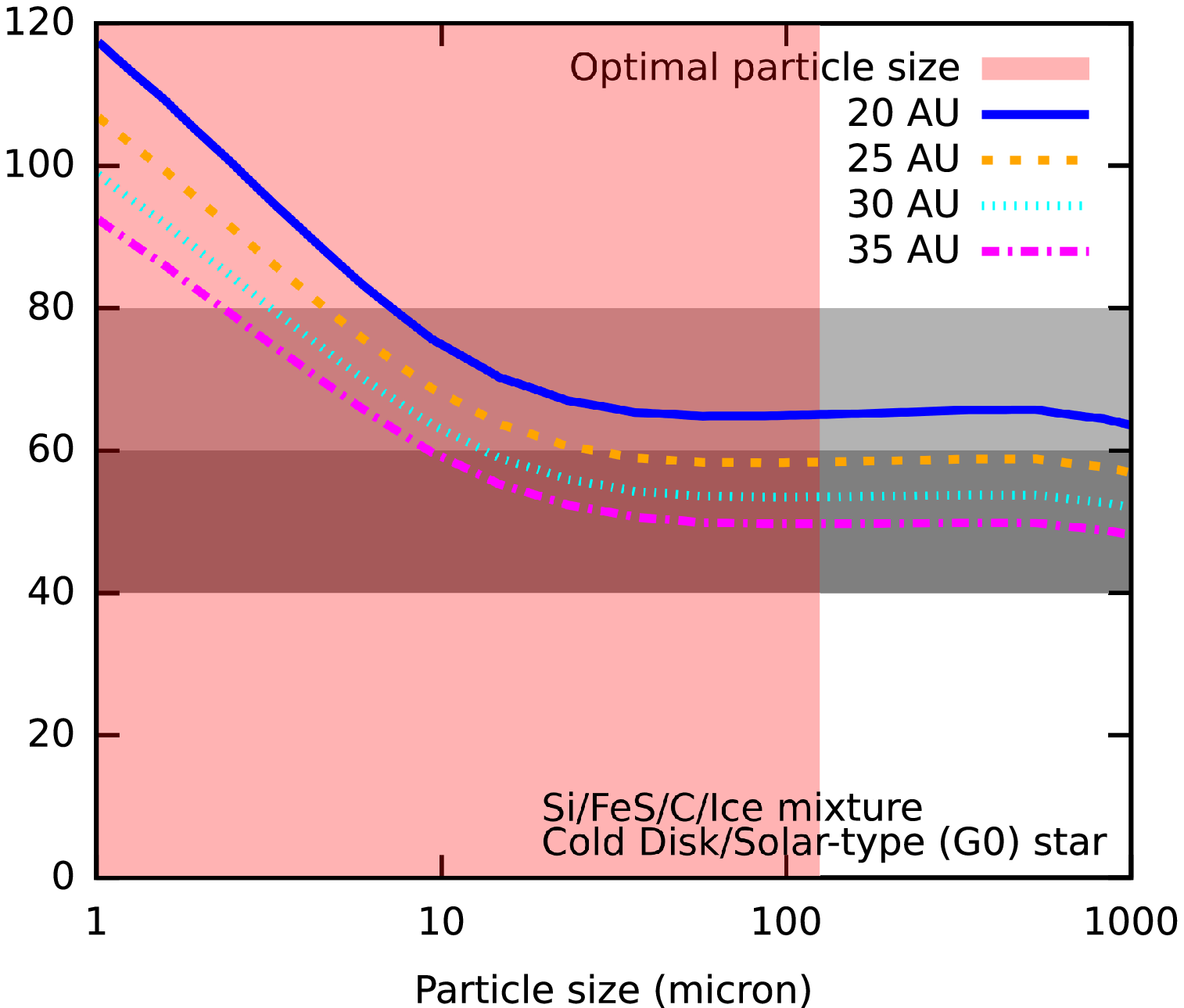}
\caption{Grain temperatures as a function of particle size, composition, radial distance, and
the spectral type of the central star. The colored vertical bands yield the optimum particle size
for emission (see text) around the certain systems, while the horizontal gray bands yield the relative number 
of systems (with 70 $\micron$ detections) found at each temperature (darker stand for more sources)
by \cite{morales11}. The plots yield the
general radial distance of warm and cold debris disks around different spectral type systems where the
colored bands, the gray bands, and the temperature curves intersect.}
\label{fig:temp}
\end{center}
\end{figure*}

The warm component was found slightly above the ice evaporation temperature, with a characteristic 
blackbody temperature of \mbox{190 K}. While the systems around solar-type stars have a narrower distribution 
in temperatures (99 to 200 K), the ones around A-type stars have a wider one 
(98 to 324 K). Assuming astronomical silicates as grain types in warm debris disks (where volatile 
elements are likely missing), we calculate the equilibrium temperatures of grains as a function of 
their sizes and radial distances around solar- and early-type stars. We show these temperature curves 
in the top panels of Figure \ref{fig:temp}. With green bands, we plot the particle size domain that 
is most effective at emitting at 24 $\micron$, when considering a realistic particle size distribution
within the disks \citep{gaspar12b}. This is found by first solving
\begin{equation}
\frac{\partial F_{24\micron}(a)}{\partial a} \equiv 0\;,
\end{equation}
and then assuming the range of particle sizes that are able to emit at or above 40\% of the peak emission
to be the effective particle size range. Since this calculation uses the modeled particle size distribution and 
realistic particle optical constants, it will differ from one system to the other.
With gray bands, we show the relative number of systems found by 
\cite{morales11} at various system temperatures. According to these plots, the most common radial 
distance for warm debris disks (where the green band and gray bands intersect) is at $\approx 3 - 6$ AU 
around solar-type (G0) stars. This can be seen in the figure because the 
temperature curves for 3.5 and 5.5 AU pass through the intersection of the green and gray bands. 
A similar argument indicates a radial distance of $\sim 11$ AU for the early-type (A0) stars. However, 
a range of distances can be accommodated, especially if one considers grains with varying optical properties.

We performed similar analysis for the cold components, but only for the solar-type sample, 
as we do not have a statistically significant sample at old ages for the early-types. For the cold component analysis, 
however, we include a second grain-type, one that includes volatiles, as these disks are located outside 
of the snowline. We use the optical properties calculated by \cite{min11} for a Si/FeS/C/Ice mixture, 
which have been used to successfully model the far-IR emission and resolved images of Fomalhaut obtained
with {\it Herschel} \citep{acke12}. We show these plots (green band - astronomical silicates; red 
band - volatile mixture), in the bottom panels of Figure \ref{fig:temp}. The plots estimate the cold 
disks to be located at around $20 - 35$ AU for an astronomical silicate composition and around $25 - 40$ AU for 
the volatile mixture. The latter estimate is more in agreement with the location of the Kuiper belt 
within our solar system. We can compare with disks around other stars by scaling their radii according 
to the thermal equilibrium distances, i.e., as $(L_{\ast})^{1/2}$. The locations for grains of the ice 
mixture generally agree with these scaled radii.

\subsection{Modeling the 24 $\micron$ excess decay}
\label{sec:24}

\begin{figure*}
\begin{center}
\includegraphics[angle=0,scale=1.4]{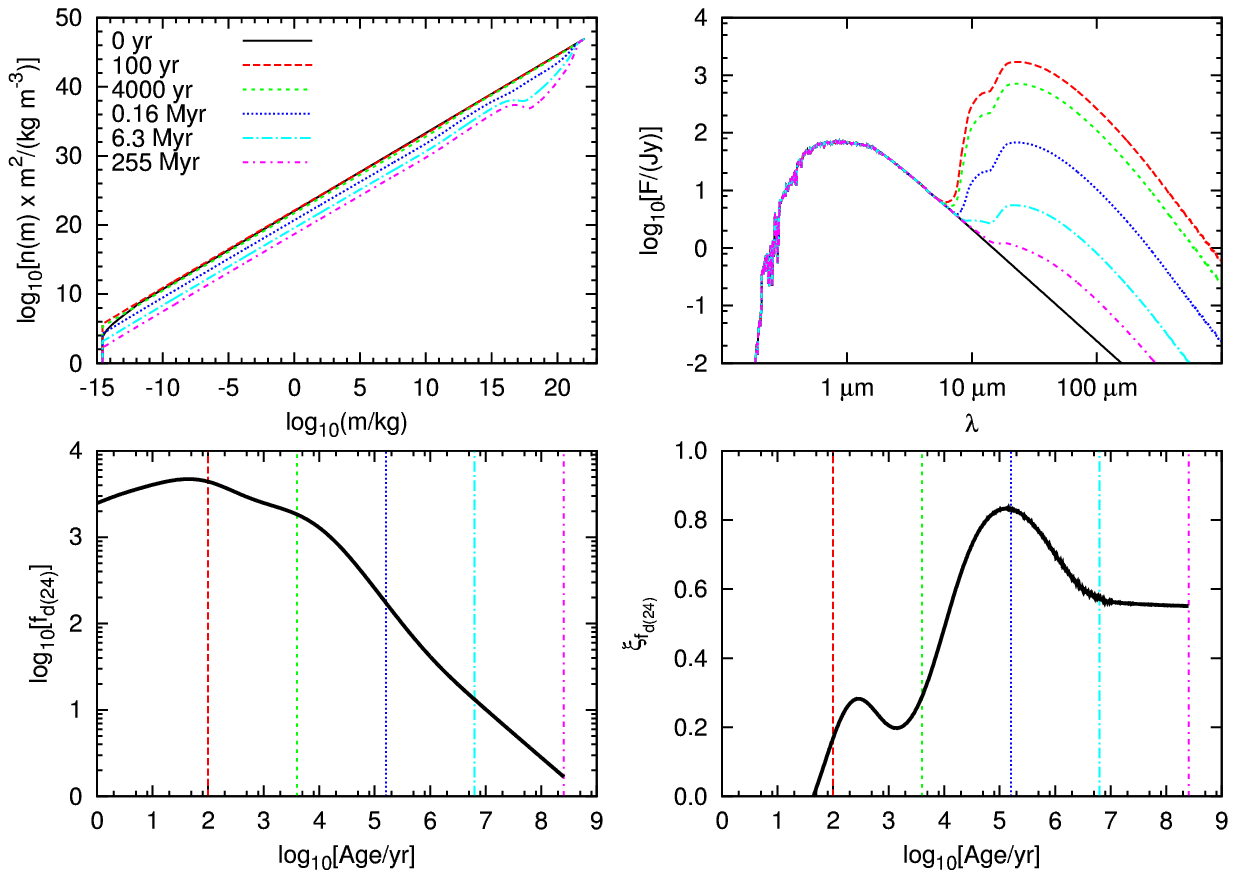}
\caption{Evolution of the warm disk component model around a solar-type star at 4.5 AU. {\it Top-left panel:}
The evolution of the particle-mass distribution in ``mass/bin''-like units. {\it Top-right panel:} Evolution
of the SED of the disk (color coding is the same as for the top-left panel). {\it Bottom-left panel:} Evolution
of the fractional 24 $\micron$ infrared emission as a function of age (the constant 5 Myr offset was applied later -- see text). {\it Bottom-right
panel:} The speed of the evolution of the fractional infrared emission. The evolution reaches its quickest point
at 0.1 Myr, and settles to a $\propto t^{-0.6}$ evolution at around 10 Myr. The average mass disk in the population, which
is barely detectable for a short period of time, would reach this at around 1000 Gyr; while a disk detectable
between 0.1 -- 1 Gyr reaches this quasi steady state around 1 Gyr.}
\label{fig:24fig1}
\end{center}
\end{figure*}

Based on the previous section, to model the decay of the warm components, we calculated the evolution 
of debris disks at radial distances between 2.5 and 10 AU with 0.5 AU increments for solar-type stars (G0), and 
at radial distances between 9.0 and 14 AU with 1.0 AU increments for early-type stars (A0). The disk widths and 
heights were set to 10\% of the disk radius, while the total disk mass was set to 100 $M_{\Earth}$, assuming a 
largest object radius of 1000 km. All other parameters were the same as for our reference model 
(Paper II). In Figure \ref{fig:24fig1}, we show the evolution of the model debris disk at 4.5 AU around 
a solar-type star. The top left panel shows the evolution of the particle mass distribution in 
``mass/bin''-like units. The top right panel shows the evolution of the SED of the debris disk, with 
the color/line coding being the same as for the mass distributions. The SEDs were calculated assuming 
astronomical silicate optical properties \citep{draine84}. Both the mass distribution and the SED 
decay steadily in the even log-spaced time intervals we picked. The bottom left panel shows the 
evolution of the fractional 24 $\micron$ infrared emission, which (as with our reference model in Section 3) 
shows varying speed in evolution. The color/line codes show the points in time that are displayed in 
the top panels. The speed of evolution is shown in the bottom right panel. The evolution speed curve 
is very similar to that of the reference model in Section \ref{sec:model}, however, the evolution is 
much quicker. While our reference model settles to the $\propto t^{-0.6}$
decay at around 100 Gyr, our warm disk model at 4.5 AU already reaches this state at 10 Myr. 
There are two reasons for this behavior: 1) The disk evolves quicker closer to the star (the reference model was 
at 25 AU), and 2) the extremely large initial disk mass (which was set to ensure coverage at large 
disk masses as well) significantly accelerates the evolution.

To compare these models with observations, we will use the excess fraction (fraction of a population 
with excesses above a threshold) as the metric, since this is the parameter most readily determined 
observationally. We calculate the fraction of sources with excesses at a given age using the 
decay of a single source and using a population synthesis routine, by making two assumptions:
\begin{enumerate}
\item The distribution of initial disk masses follows a log--normal function.
\item All systems initiate their collisional cascade at the same point in time during their evolution. This point can not
be earlier than the time of planet formation. We fix $t(0)$ at 5 Myr for our calculations.
\end{enumerate}
Both assumptions are plausible. Our first assumption is 
consistent with observations of protoplanetary disks, as shown by \cite{andrews05}. In addition, this form was 
adopted by \cite{wyatt07} as the starting point for their analytic modeling of debris disk evolution, and thus adopting 
a similar initial form allows direct comparisons with this previous work. The log-normal form also gives a reasonably good 
fit to the distribution of excesses in young debris systems (J.\ Sierchio et al., Submitted to ApJ). 
We define the probability density distribution of the total disk masses as
\begin{equation}
n (M_{\rm tot}; \mu, \sigma_e) = \frac{1}{M_{\rm tot}\sqrt{2\pi\sigma_e^2}}{\rm Exp}\left\{-\frac{\left[\ln\left(M_{\rm tot}\right) -
\mu\right]^2}{2\sigma_e^2}\right\}\;,
\end{equation}
where $n(M_{\rm tot})$ is the probability density of systems with initial masses of $M_{\rm tot}$, 
the ``location parameter'' of the log--normal 
distribution is $\mu$, and $\sigma_e$ is the ``scale parameter''. We set the scale parameter to be equal to the
width of the distribution of protoplanetary disk masses found by \cite{andrews05}, $\sigma^2_{\rm e} = 6.95\pm0.06$ (in natural log base). 
Since the peak in the mass distribution depends on the largest mass within the systems and can be arbitrarily varied to a large extent,
the location parameter is found by fitting. We set the median (geometric mean) of our log--normal distribution of masses to be equal to
\begin{equation}
C M_{\rm tot,0} = {\rm e}^{\mu}
\end{equation}
where $C$ is a scaling constant that yields the scaling offset between the median mass of the distribution and the 
mass of our reference model ($M_{\rm tot,0}=100 M_{\Earth}$). 

The second assumption arises because the collisional cascades in debris disks cannot be maintained without larger 
planetary bodies shepherding and exciting the system. According to core accretion models, giant planets 
such as Jupiter and Saturn form in less than 10 Myr \citep{pollack96,ida04}, while disk instability models 
predict even shorter timescales \citep{boss97,boss01}. As planets form, simultaneously, the protoplanetary 
disks fade \citep{haisch01}, and their remnants transition into cascading disk structures. Based on these
arguments, our $t(0)$ value of $5~{\rm Myr}$ is reasonable. Our assumption ignores the possibility of 
later-generation debris disks. That is, any late-phase dynamical activity that yields substantial amounts 
of debris will not be captured in our model, whose assumptions are similar to those of the \cite{wyatt07} analytic 
model in which the disk evolution is purely decay from the initial log-normal distribution.

A useful property of collisional models is that their evolution scales according to initial mass, which made the synthesis
significantly simpler, as only a single model had to be calculated. The flux $f$ emitted by a model at time
$t$ with an initial mass $M_{\rm tot}$ will be equal to a fiducial model's flux 
$f_0$ with initial mass $M_{\rm tot,0}$ at time $t_0$ as
\begin{equation}
f (t) = f_0 (t_0) \frac{M_{\rm tot}}{M_{\rm tot,0}}
\end{equation}
\begin{equation}
t = t_0 \frac{M_{\rm tot,0}}{M_{\rm tot}}\;.
\end{equation}
We verified that our model follows these scaling laws by running multiple models with varying initial disk masses (see Appendix).
These relations are equivalent to a translation of the decay along a $t^{-1}$ slope, which is why as long as the decay
of single sources remains slower than $t^{-1}$, the decay curves will not cross each other. Similar behavior has been 
shown by \cite{lohne08}. This also means that each particular observed $f(t)$ value can be attributed to a particular 
initial disk mass and that at any given age the limiting mass can be calculated that yields a fractional infrared 
emission that is above our detection threshold.

To compare with the observationally determined percentage of sources above a given detection threshold, we need to
find the initial mass whose theoretical decay curve yields an excess above this threshold as a function of system age.
As detailed above, since the decay speed is always slower than $t^{-1}$, this will always be a single mass limit, 
without additional mass ranges. We can then calculate the cumulative distribution function (CDF) of the log--normal 
function using these initial mass limits [$M_l(t)$] defined as
\begin{equation}
{\rm CDF}\left[M_l(t); \mu, \sigma_e\right] = \frac{1}{2}\left(1+{\rm erf}\left\{\frac{\ln\left[M_l(t)\right]-\mu}{\sqrt{2\sigma_e^2}}\right\}\right)\;.
\end{equation}
Although the distributions get skewed in the number density vs.\ current mass (or fractional infrared emission) vs.\ age 
phase space, they remain log--normal in the number density vs.\ initial mass phase space, which is why this method 
can be used. The $\chi_{\rm fit}^2$ of our fitting procedure, where we only fit the location of the peak of the mass 
distribution, is then
\begin{equation}
\chi_{\rm fit}^2 = \sum_i\frac{\left\{1-{\rm CDF}\left[M_l(t_i);\mu,\sigma_e\right] - F\left(t_i\right)\right\}^2}{\sigma_F^2\left(t_i\right)}\;,
\end{equation}
where $F\left(t_i\right)$ is the measured excess fraction at time $t_i$, and $\sigma_F^2\left(t_i\right)$ is the error 
of the measured excess fraction at time $t_i$. It is necessary to subtract the CDF from 1, because we are comparing
the percentage of sources above our threshold and not below.

\begin{figure*}
\begin{center}
\includegraphics[angle=0,scale=1.3]{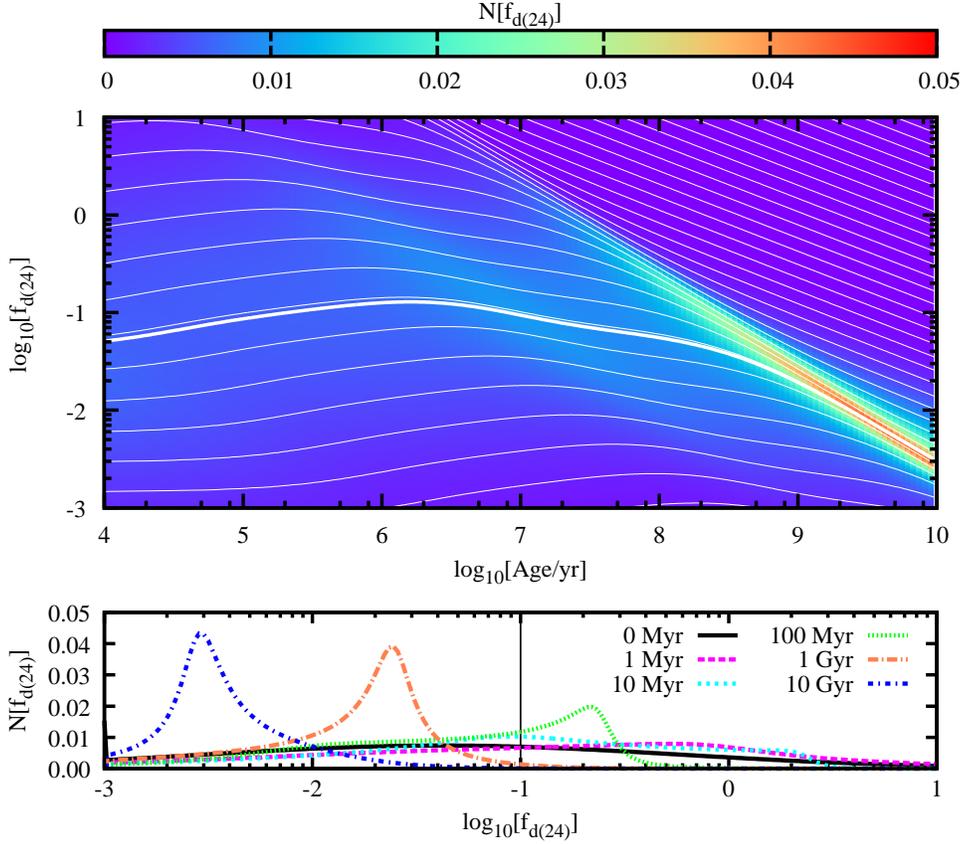}
\caption{The best fitting mass population and its evolution for the warm component
of solar-type stars placed at 4.5 AU. {\it Top panel:} the fractional 24 $\micron$ emission decay curves ($f_{d(24)} = F_{\rm disk}(24)/F_{\ast}(24)$), shifted along
the $t^{-1}$ slope as a function of varying initial disk masses. The color code of the plot is proportional
the number of systems at any given point in the phase space. The bold line represents the evolution of the average mass
disk in the population. {\it Bottom panel:} the evolution of the 
number distribution as a function of fractional 24 $\micron$ infrared emission 
at different ages (vertical cuts along the top panel).
The initial fractional infrared emission distribution at age 0 follows the initial mass distribution's log--normal 
function, however, as the population evolves this gets significantly skewed. The black vertical line at 
$f_{d(24)} = 0.1$ gives our detection threshold at $24~\micron$ and the lower integration limit for our 
excess fraction decay calculations}
\label{fig:popsynth}
\end{center}
\end{figure*}

In Figure \ref{fig:popsynth}, we show the best fitting mass population and its evolution for the warm component
of solar-type stars placed at 4.5 AU. The top panel shows the fractional infrared emission decay curves, shifted along
the $t^{-1}$ slope as a function of varying initial disk masses. As the plot shows, the curves do not intersect,
and they do not reach a common decay envelope (as is predicted by analytic models that yield a uniform $t^{-1}$
decay slope \cite[e.g.][]{wyatt07}). The decay curves do merge after 500 Myr of evolution, leaving
a largely unpopulated (but not empty) area in the upper-right corner of the plot. Before 500 Myr, they
occupy most of the phase space. For cold debris disks, the merging of the decay curves happens at an even later
point in time. This also means that a maximum possible disk mass or fractional infrared emission 
at a given age, as predicted by the simple analytic models, does not exist; although with
adjustments to the slower decay, after 500 Myr, they could approximate the evolution of the population. The 
color code of the plot shows the number of 
systems at any given point in the phase space. While systems show a spread in fractional infrared emission
up to $\approx 100~{\rm Myr}$, after that they do increase in density along the decay curve of the average disk mass 
(shown with bold line) up to 10 Gyr, which is still faster ($\propto t^{-0.6}$ \dots $t^{-0.8}$) than the final 
quasi steady state decay speed of $\propto t^{-0.6}$. The bottom panel shows the evolution of the number 
distribution as a function of fractional 24 $\micron$ emission at different ages 
(vertical cuts along the top panel). The initial distribution at age 0 follows the
initial mass distribution's log--normal function; however, as the population evolves this gets significantly 
skewed. The black vertical line at $f_{d(24)} = 0.1$ gives our detection threshold at $24~\micron$ and the lower
integration limit for our excess fraction decay calculations.

\begin{figure}
\begin{center}
\includegraphics[angle=0,scale=0.7]{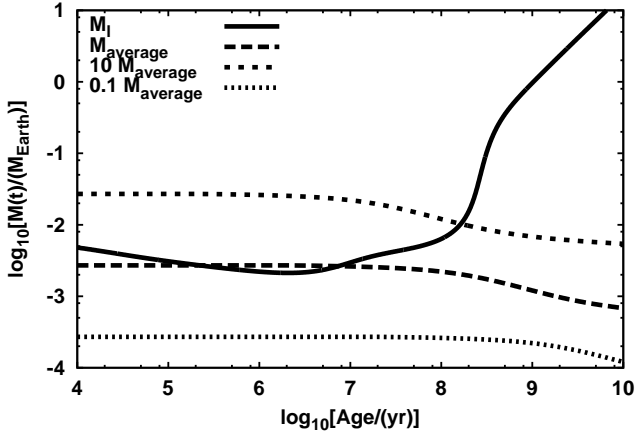}
\caption{The mass of the disk at the detection limit as a function of system age [$M_l(t)$],
and the evolution of the average disk mass ($\pm$ 1 dex) in the distribution.}
\label{fig:ml}
\end{center}
\end{figure}

Figure \ref{fig:ml} shows the calculated $M_l(t)$ limit as a function of system age as well as the average mass
of our modeled population ($\pm 1$ dex). The plot shows that any system with excess that is over a Gyr old could only
be explained with the quasi steady state model if its initial mass was at least 3 - 4 orders of magnitude larger than
the mass of our average disk. Since such massive disks are unlikely, these late phase excesses must arise from either
a stochastic event or possibly from small grains leaking inward from activity in the outer cold ring.

\begin{figure*}
\begin{center}
\includegraphics[angle=0,scale=0.54]{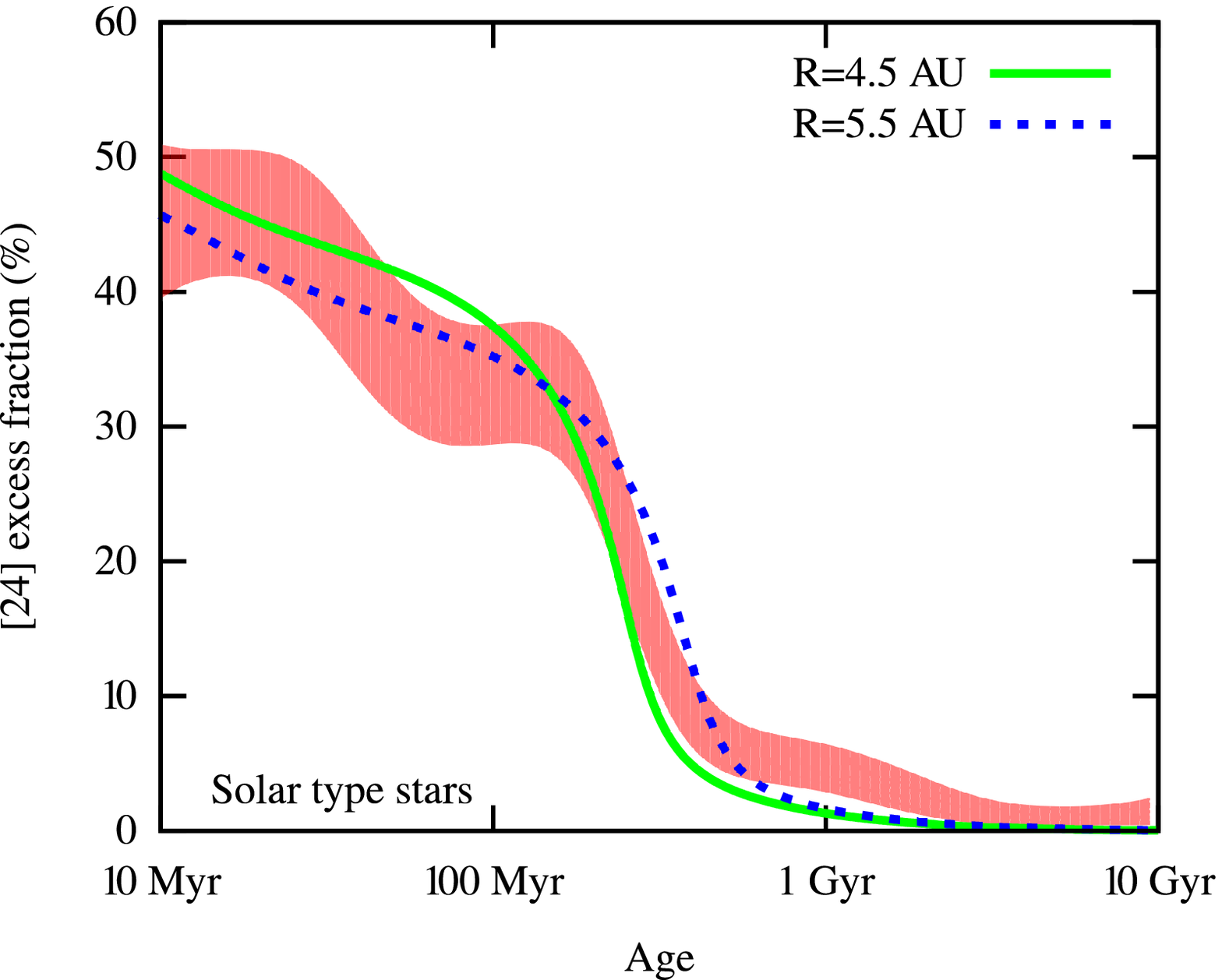}
\includegraphics[angle=0,scale=0.54]{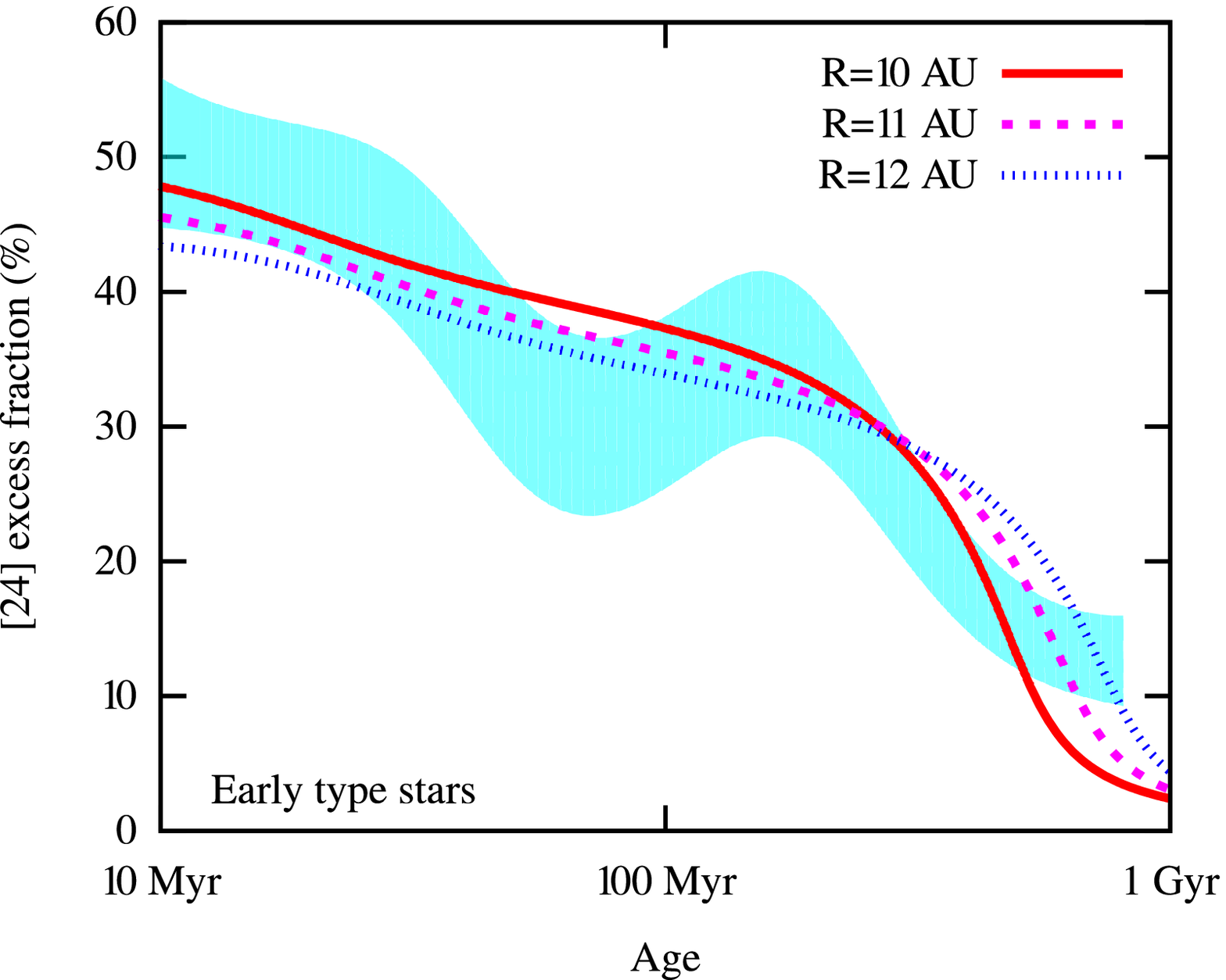}
\caption{The excess fraction decay curves calculated from our best fitting population
synthesis models for warm disks at varying distances for the two different spectral groups. The {\it right panel} shows the 
models for the solar-type stars, while the {\it left panel} shows them for the early-type stars.}
\label{fig:decays}
\end{center}
\end{figure*}

In Figure \ref{fig:decays}, we show the excess fraction decay curves calculated from our best fitting population
synthesis models at varying distances for the two different spectral groups. The {\it left panel} shows the 
models for the solar-type stars, while the {\it right panel} shows them for the early-type stars. The
solar-types can be adequately fit with models at 4.5 and 5.5 AU, which matches reasonably well to the temperature peak observed 
by \cite{morales11}. Similarly, we get adequate fits to the early-type population with models placed at 11 AU, which 
is also in agreement with the temperature peak observed by \cite{morales11} and our radial distance constraint.

Our population synthesis routine yields excess fraction decays that are in agreement with the observations. This
is the first time that a numerical collisional cascade code has been used together with a population synthesis
routine to show agreement between the modeled and the observed decay of infrared excess emission originating 
from debris disks. The average initial disk mass predicted by our population synthesis has a total of 
0.23 $M_{\rm Moon}$, with
a largest body radius of $1000~{\rm km}$. This yields dust masses of 
$M_{\rm dust} (<1~{\rm cm}) = 2.3\times10^{-5}~M_{\rm Moon}$ $= 2.8\times10^{-7}~M_{\Earth}$ ($M_{\rm dust} (<1~{\rm mm}) = 7.3\times10^{-6}~M_{\rm Moon}=9.0\times10^{-8}~M_{\Earth}$). Our predicted average dust mass is in agreement
with the range of dust masses ($2.8\times10^{-7}$ to $5.2\times10^{-3}~M_{\rm Moon}$) observed by \cite{plavchan09}
for debris disks around young low-mass stars, determined from infrared luminosities.

\subsection{Modeling the far-IR (70--100 $\micron$) excess decay}
\label{sec:farIR}

According to section \ref{sec:location}, to model the decay of the cold disks, we calculated the evolution 
of a disk placed at 15, 20, 25, 30, and 35 AU around a solar-type star. At these distances, volatiles are a 
large part of the composition, which will change not only the optical properties of the smallest grains (see
section \ref{sec:location}), but also the tensile strength of the material. To account for this,
we used the tensile strength properties of water-ice from \cite{benz99} and the erosive cratering properties
of ice from \cite{koschny01a,koschny01b}. For comparison, we repeated the calculations with the tensile strengths 
of basalt, as in our reference model. The emission of the modeled particle size distributions was calculated assuming
astronomical silicates for the regular basalt tensile strength models, and the volatile mixture \citep{min11}
mentioned in section \ref{sec:location} for the water-ice tensile strength models.

Understanding and modeling the decay observed at far-IR wavelengths is significantly more difficult than it is
for its shorter, 24 $\micron$ wavelength, counterpart. This is due to the non-uniform detection limits 
at longer wavelengths, which are frequently significantly above the stellar photospheric values. Here, we will 
use the method developed by Sierchio et al.\ (Submitted tp ApJ) to study the evolution of the far-IR excess, but slightly 
modified to use our calculated evolved fractional infrared emission distributions. This new method quantifies the decay,
taking into account both detections and non-detections and also the non-uniform detection limits.

We define the significance of an observed excess as
\begin{equation}
\chi = \frac{F - P}{\sigma} = \frac{R_f - 1}{\sigma_R}\;,
\end{equation}
where $F$ is the detected flux, $P$ is the predicted photospheric emission of
the central star, while $\sigma$ is the error of the photometry. We define $R_f = F/P$ as
the excess ratio of the source, and $\sigma_R$ as the photosphere normalized error. 

The majority of the sources had both {\it Spitzer} 70 $\micron$ and {\it Herschel} PACS 100 $\micron$ data. 
We merged these data to simulate a single dummy 85 $\micron$ datapoint as
\begin{equation}
R_{f85} = \frac{R_{f70}/\sigma_{70}^2 + R_{f100}/\sigma_{100}^2}{1/\sigma_{70}^2 + 1/\sigma_{100}^2}\;,
\end{equation}
with an error of
\begin{equation}
\sigma_{R_{f85}} = \frac{1}{\left(1/\sigma_{70}^2 + 1/\sigma_{100}^2\right)^{1/2}}\;.
\end{equation}
Since the excess ratios at 70 and 100 $\micron$ are similar, when measurement was only available at a single band, 
it was assigned to be at 85 $\micron$. As discussed in 
Sections \ref{sec:MIPS70} and \ref{sec:PACS100}, the definitions of excesses at the far-IR wavelengths are 
determined on a case-by-case basis for the detected disks. For the modeling comparison,
a $\chi$ limit is required however, defining an excess. We chose $\chi_{85} >= 3.7$ as our detection threshold,
which recovers 63 of the 66 excess sources and adds only 2 false identifications.

We separate our observed sources into three age bins that cover the age range between 0 and 10 Gyr, the first bin
including stars up to 1 Gyr (median age of sources: 475 Myr), the second including stars with ages between 
1 and 4 Gyr (median age of sources: 2.65 Gyr), and the third with stars between 4 and 10 Gyr (median age 
of sources: 6.54 Gyr). These age bins were chosen to include equal numbers of sources (143,143, and 144, respectively).

We synthesize disk populations at 85 $\micron$ the same way as we did when modeling the $24~\micron$ excess decay, 
assuming a log--normal initial mass distribution, with the scale parameter fixed at $\sigma_e^2=6.95$, and varying only the
location parameter of the distribution.

Finally, we compare the calculated distribution at 475 Myr, 2.65 Gyr, and at 6.54 Gyr, to the observed first, second, 
and third data bins, respectively. Since the detection thresholds are non-uniform, instead of doing a straight 
comparison between the distributions, we calculate the number of possible detections from
our modeled distributions and compare with the observed distribution of excess significances ($\chi$'s). 
Assuming that the model distribution does show the underlying distribution of fractional
far-IR excesses, we integrate the distribution upward from the detection threshold for each star in the corresponding data
bin. The detection threshold is given as
\begin{equation}
\Theta = 1 + 3\frac{\sigma}{P} = 1 + 3\frac{R_f - 1}{\chi} = 1 + 3\sigma_R\;.
\end{equation}
Integrating the distribution from the respective detection threshold of each source yields the probability of detecting
an excess at the given threshold according to the model. Summing up these probabilities then yields the total number of 
predicted excesses that would be detected. This can then be compared to the actual number of observed excesses. The model that 
yields the best agreement for all three data bins consistently is defined as the best fitting model.

\begin{figure}
\begin{center}
\includegraphics[angle=0,scale=0.48]{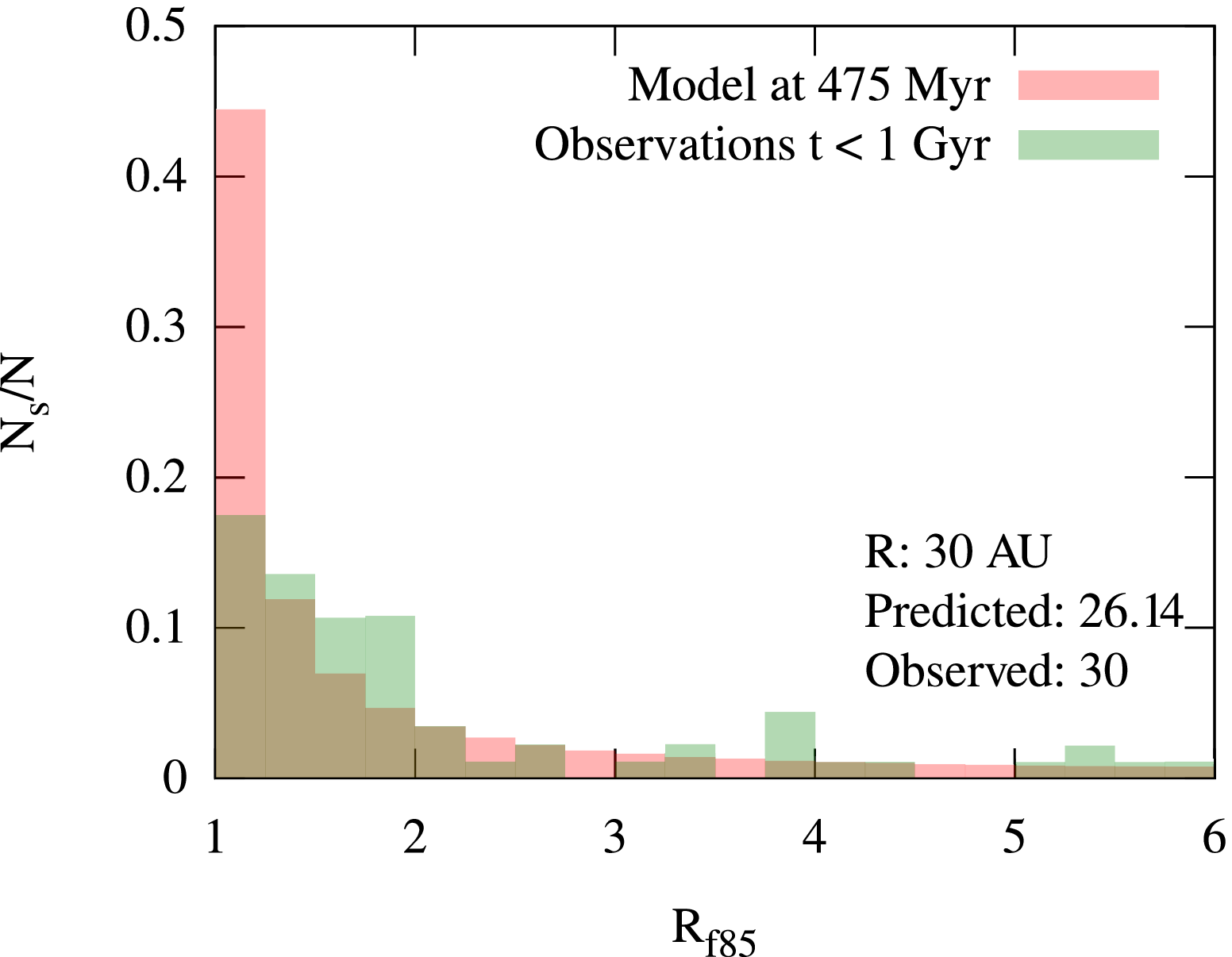}

\includegraphics[angle=0,scale=0.48]{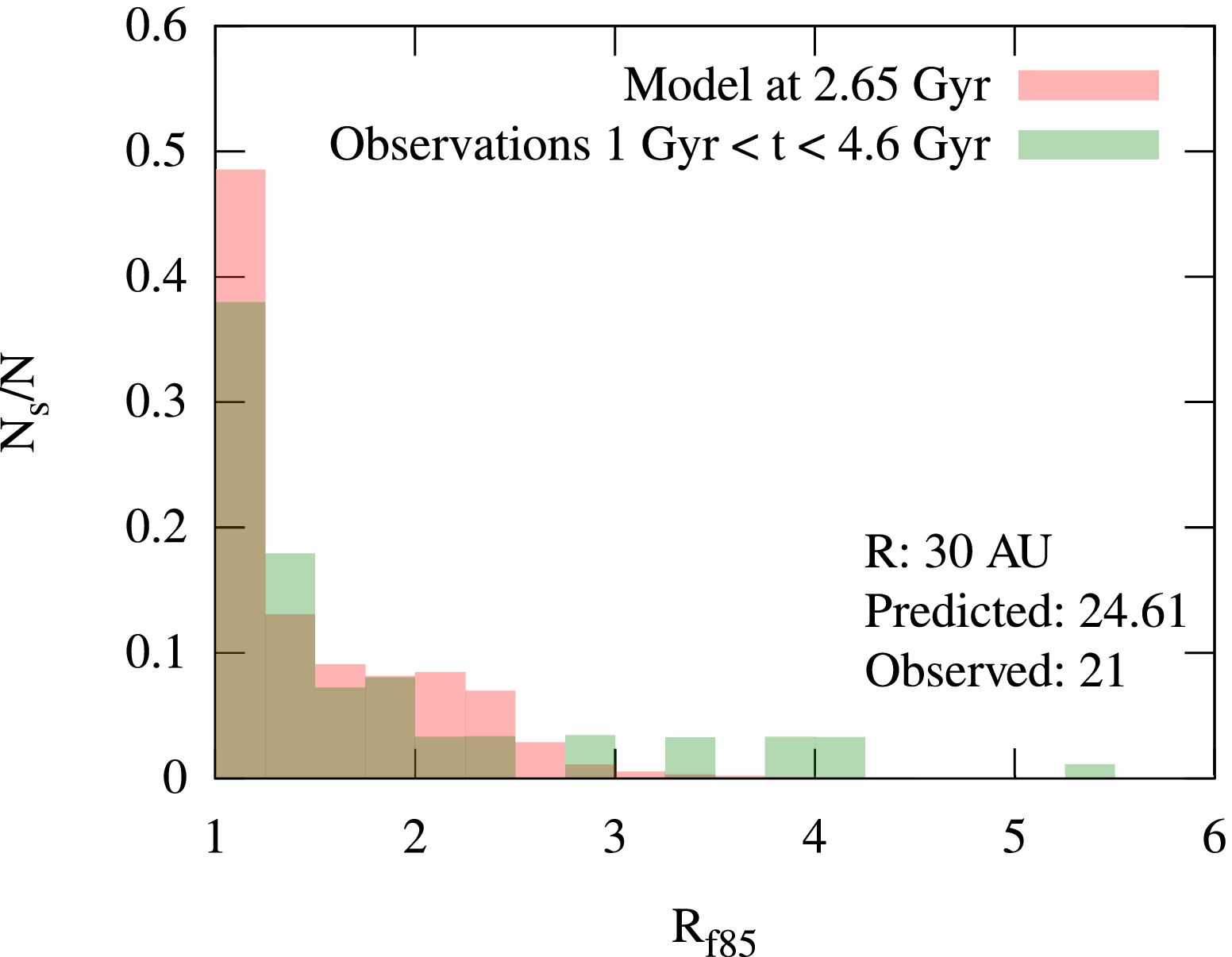}

\includegraphics[angle=0,scale=0.48]{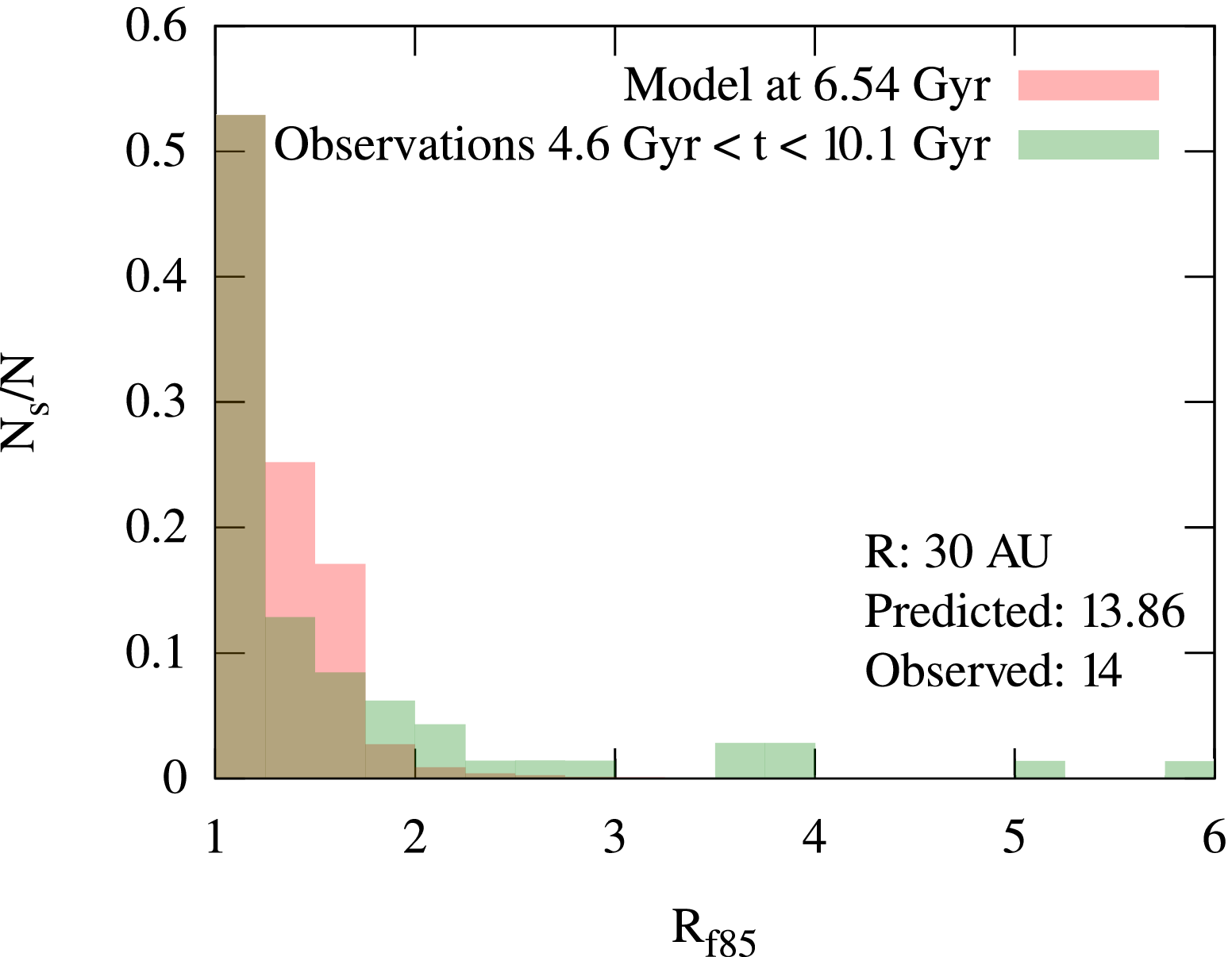}
\caption{The observed and modeled distribution of excesses at 30 AU around a solar-type star, using water-ice 
particle tensile strength and a volatile mix for grain optical properties. The best fitting model using the fiducial
basalt tensile strength and astronomical silicates for grain properties yielded similar distributions, only at 17.5 AU
radial distance.}
\label{fig:R70}
\end{center}
\end{figure}

\begin{figure}
\begin{center}
\includegraphics[angle=0,scale=0.5]{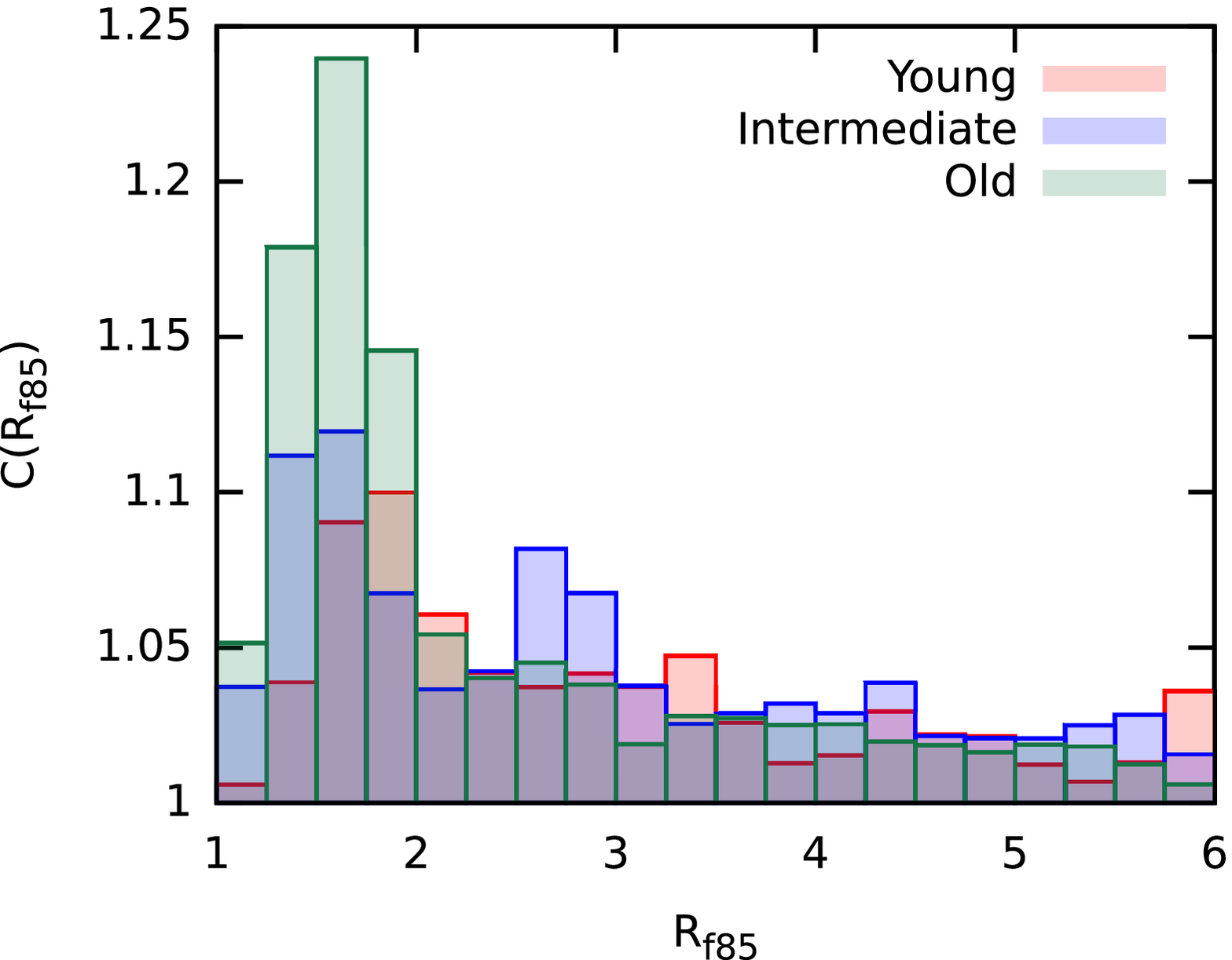}
\caption{The calculated completeness correction for the PACS 100 $\micron$ data. See text for details.}
\label{fig:CC}
\end{center}
\end{figure}

In Figure \ref{fig:R70}, we show the observed and modeled distribution of excesses at 30 AU, assuming
water-ice tensile strength and the ice-mixture optical properties (the best fitting solution) in the 
three separate age bins. The observed sources are completeness corrected and sources below 
$R_f < 1$ are not shown. For completeness correction, we 
assumed that the observed data well represents the photometric error distribution 
$\Gamma(\sigma_R)$ of PACS observations. Then for each $\Delta R_f$ bin, we determined the probability of a source 
being in that bin, assuming the previously defined error distribution, yielding,
\begin{equation}
N_s(R_f) = \int_{\sigma_{R-}}^{\sigma_{R+}} \Gamma(\sigma_R) d\sigma_R\;,
\end{equation}
where
\begin{eqnarray}
\sigma_{R-} &=& \frac{R_- - 1}{\chi_{\rm det}} \\
\sigma_{R+} &=& \frac{R_+ - 1}{\chi_{\rm det}}\;.
\end{eqnarray}
Here, $R_-$ and $R_+$ represent the lower and upper boundaries of the $\Delta R_f$ bin, respectively,
as before $\sigma_R$ is the photospheric flux normalized error, and $\chi_{\rm det}$ is the detection
threshold of $\chi$. We adopt $\chi_{\rm det} = 3.7$ based on our data. The completeness correction 
than can be calculated as
\begin{equation}
C(R_f) = \frac{N}{N-N_s(R_f)}\;,
\end{equation}
where $N$ is the total number of sources. In Figure \ref{fig:CC}, we show the completeness correction 
curve we derived for the combined DEBRIS and DUNES surveys at 100 $\micron$ for the three age groups we analyzed.

\begin{figure*}
\begin{center}
\includegraphics[angle=0,scale=0.7]{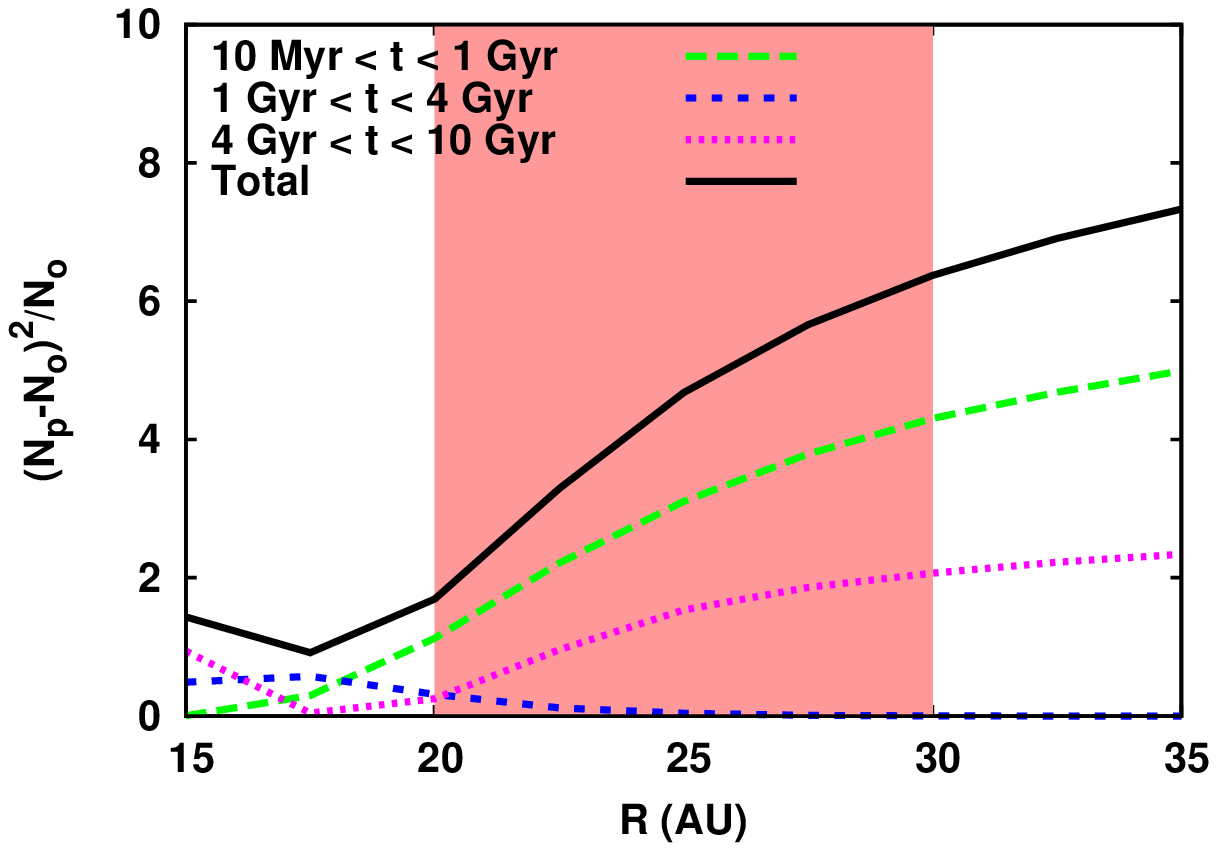}
\includegraphics[angle=0,scale=0.7]{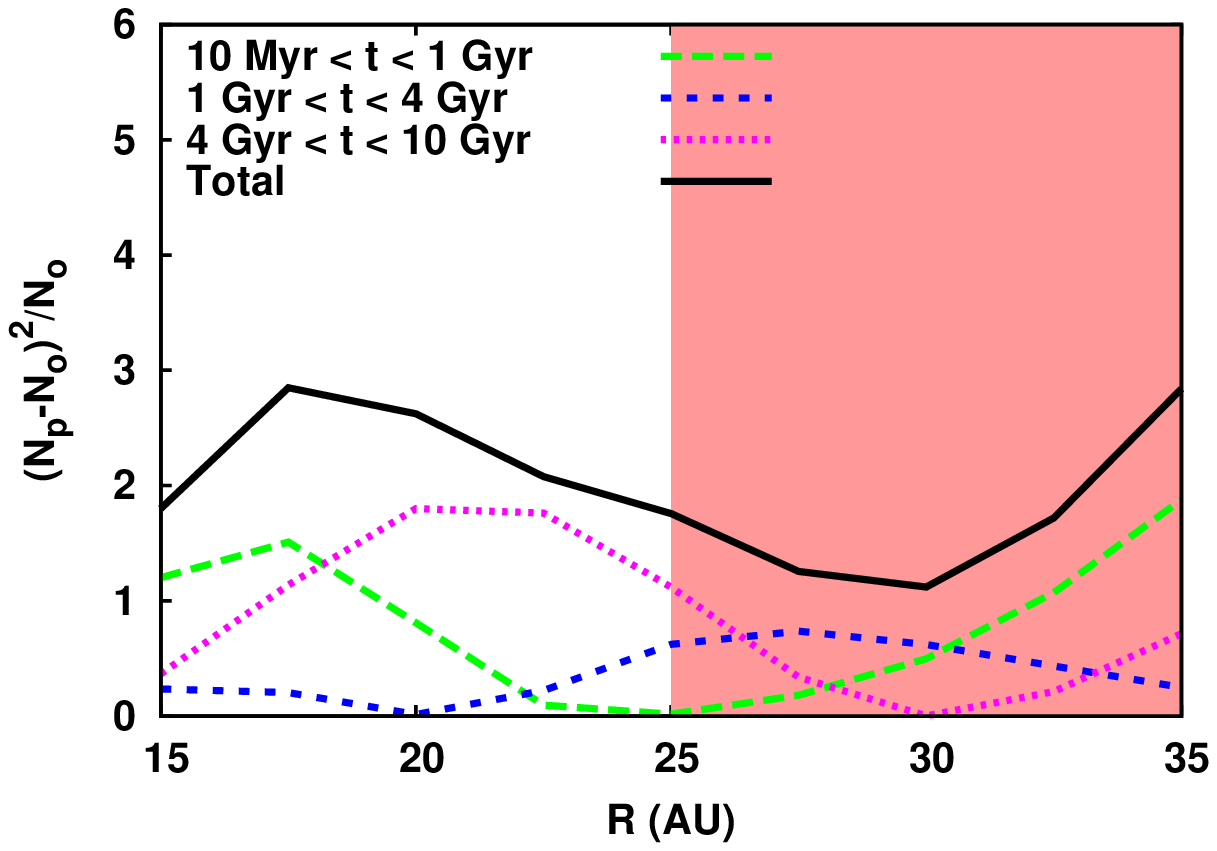}
\caption{The relative difference between the predicted and observed number of far-IR excess sources at solar-type stars,
as a function of model radial distance. {\it Left panel:} Basalt tensile strength and astronomical silicate grain 
optical properties. {\it Right panel:} Water-ice tensile strength and volatile mixture grain optical properties.}
\label{fig:70sol}
\end{center}
\end{figure*}

The panels in Figure \ref{fig:R70} display the number of sources observed and predicted by our calculations 
in each given age bin. We emphasize, that the numbers of predicted sources are {\bf not} determined based on these
 binned emission plots, but with the method detailed
above. These plots show the emission distribution predicted by our fits and compare it with the
completeness corrected observed distributions. The distributions are scaled to the total number of sources. 
The best fit for the basalt tensile strength and astronomical silicate optical property model (which looks
almost identical to the ice mixture/strength solution plotted) was at $\approx 17.5 ~{\rm AU}$, which is clearly 
inwards of the predictions we made in section \ref{sec:location}, and inwards of the cold
disk component of our solar system. However, the water-ice composition and tensile strength model yields
a fit at 30 AU, which is in agreement with the predictions and with the
placement of the inner edge of the Kuiper belt in our solar system. In Table \ref{tab:70}, we tabulate the number of predicted 
and observed sources for both models at various radial distances, and in Figure \ref{fig:70sol} we plot the 
relative differences between these numbers and show the predicted radial location of the disks with a red band.
In Table \ref{tab:70}, we also give the median masses of the best fitting distributions for each model.
For our best fitting model (ice mixture particles at 30 AU), the median initial mass of the distribution is $0.028~M_{\Earth}$,
with a surface density of $1.3\times10^{-3}~{\rm g~cm}^{-2}$, which is over four orders of magnitude times underdense 
compared to the minimum-mass-solar-nebula surface density.

\setcounter{table}{4}
\begin{center}
\begin{deluxetable*}{l|cccc|cccc}
\def\arraystrech{4}
\tablecolumns{9}
\tablewidth{0pt}
\tablecaption{The number of cold debris disk sources around solar-type stars predicted vs.\ the number of 
sources observed in each age bin at different disk radii , assuming varying optical properties and particle
tensile strengths. The predicted radial location of the disks is between 20 to 35 AU for the silicate composition and 25 to 40 AU
for the ice mixture. The median mass of the best fitting distribution
is also given, assuming a largest body with a radius of 1000 km. \label{tab:70}}
\tablehead{
\colhead{} 		& \multicolumn{4}{c}{$N_{P}/N_{O}$ for Silicates [$Q_{D}^{\ast}$(Basalt)]} & \multicolumn{4}{c}{$N_{P}/N_{O}$ for Si/FeS/C/Ice Mixture [$Q_{D}^{\ast}$(Ice)]}\\
\colhead{R} 		& \colhead{$M_{\rm med}$}  &	\colhead{0.01 \dots 1} & \colhead{1 \dots 4} & \colhead{4 \dots 10} & \colhead{$M_{\rm med}$}  & \colhead{0.01 \dots 1} & \colhead{1 \dots 4} & \colhead{4 \dots 10}\\
\colhead{(AU)} 		& \colhead{($M_{\Earth}$)} &    \colhead{(Gyr)}        & \colhead{(Gyr)}     & \colhead{(Gyr)}      & \colhead{($M_{\Earth}$)} & \colhead{(Gyr)}        &\colhead{(Gyr)}      & \colhead{(Gyr)}
}
\startdata
15 & 0.051 & 29.61/30 & 24.21/21 & 10.37/14 & 0.397 & 36.00/30 & 18.78/21 & 11.74/14 \\
20 & 0.029 & 24.19/30 & 23.57/21 & 15.88/14 & 0.092 & 34.93/30 & 20.48/21 & 8.98/14  \\
25 & 0.023 & 20.35/30 & 21.92/21 & 18.63/14 & 0.039 & 29.28/30 & 24.61/21 & 10.05/14 \\
30 & 0.024 & 18.64/30 & 21.18/21 & 19.38/14 & 0.028 & 26.14/30 & 24.61/21 & 13.86/14 \\
35 & 0.026 & 17.76/30 & 20.85/21 & 19.72/14 & 0.022 & 22.50/30 & 23.29/21 & 17.16/14 
\enddata
\end{deluxetable*}
\end{center}

\subsection{Disk incidence for old stars}

At 24 $\micron$, our model suggests there should be virtually no detected debris disks around stars older than 
1 Gyr. Nonetheless, there are a number of examples, and examination of their ages indicates that they are of 
high weight. This result implies that the simple assumption \citep[e.g.,][]{wyatt07} that debris disks can be
modeled consistently starting from a log-normal initial mass distribution is successful up to about a Gyr, but
that there are additional systems around older stars above the predictions of the simple model. We attribute
these systems in part to late-phase dynamical activity that has led to substantial enhancements in dust production. 
Two examples in our sample are HD 69830 \citep{beichman05} 
and $\eta$ Crv \citep{lisse12}. Another example is BD+20 307 \citep{song05}. All three of these systems have strong 
features in their infrared spectra that indicate the emission is dominated by small grains that must be recently produced, 
which supports the hypothesis that they are the sites of recent major collisional events. These systems
with late phase 24 $\micron$ excess, however, could also be explained by grains leaking inward from an active cold ring.

Similarly, although our model successfully matches the numbers of detected disks in the far infrared, the 
observations find many more large excesses than predicted (Figure \ref{fig:R70}, bottom panel). A plausible 
explanation would be that the outer, cold disk component can also have renaissance of dust production due to 
late phase dynamical activity.

\section{Constraining model parameters with observations}
\label{sec:constr}

\begin{figure*}
\begin{center}
\includegraphics[angle=0,scale=0.75]{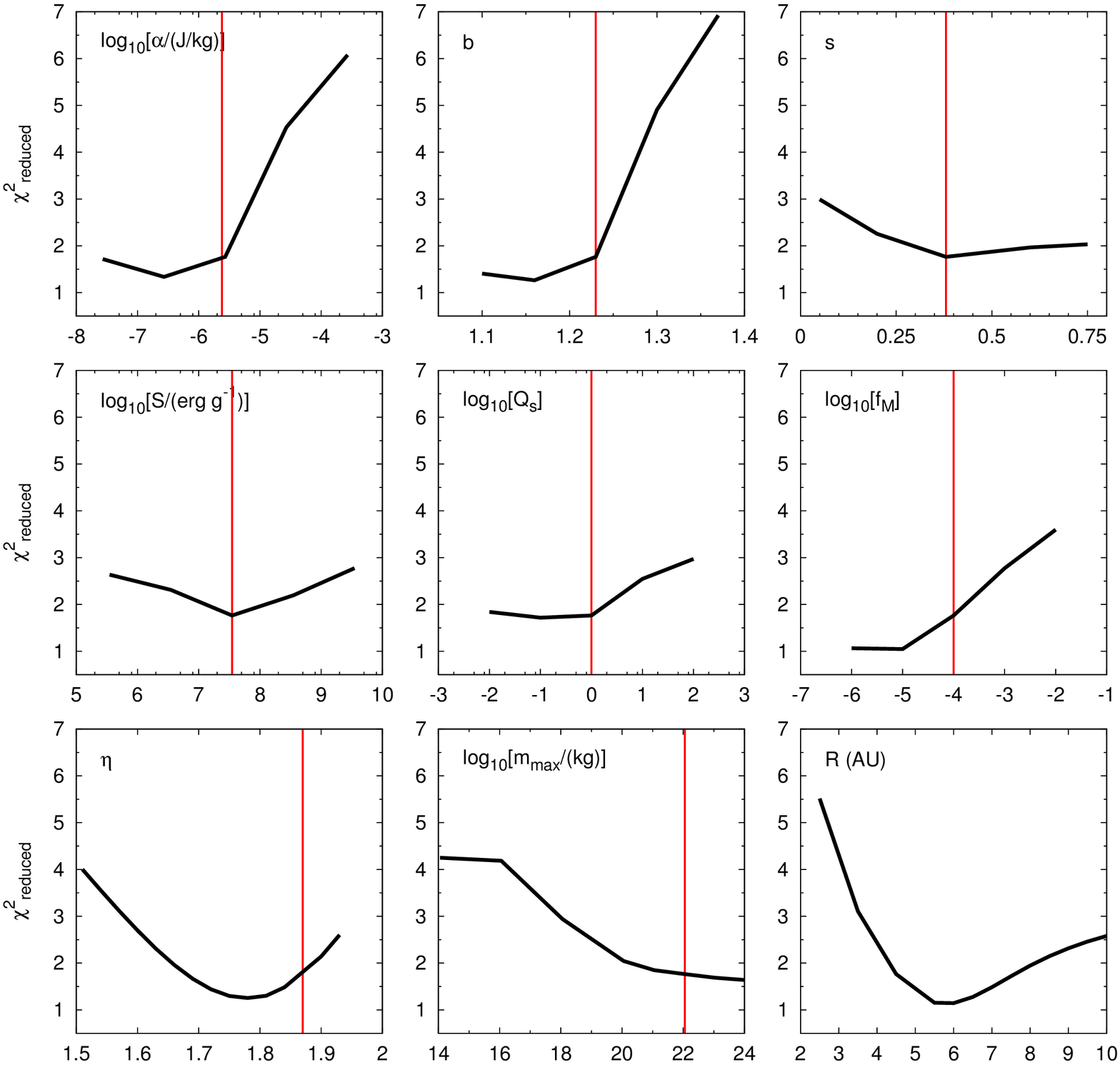}
\caption{The reduced $\chi^2$ minima of the model population fits for each tested value of the selected 
nine model variables that have the largest effects within the fits. Red lines show the values of variables 
used in the base model (the warm debris disk around a solar-type star at 4.5 AU introduced in Section \ref{sec:24}).}
\label{fig:chis}
\end{center}
\end{figure*}

We ran more than a hundred extra models, taking our best fit to the decay of the warm component of solar-type debris disks
at 4.5 AU as the basis, to test the dependence of the decay on the variables of the model. We varied each model parameter
within a range of values and performed the same population synthesis routine and fitting as we did in Section 4.
Of these, nine variables show signs of having some effect on the evolution of the excess fraction decay curve. 
In Figure \ref{fig:chis}, we present the reduced $\chi^2$ minima at each value of these nine parameters.

{\it Variables} $\alpha$ {\it and} $b$ of the cratered mass equation had the strongest effect on the slope of the evolution 
(see Appendix) and also strongly affect the population synthesis fits. Values of $\alpha$ and $b$ that describe 
materials that are softer in erosive collisions ($\alpha > 10^{-5}$ J kg$^{-1}$, $b > 1.27$) can be generally ruled out by our analysis
for the warm component of debris disks. Our analysis also shows that the measured values of these variables, which we 
used in our reference models, yield acceptable fits with our population synthesis routine. This is similar to the effect 
we observed when using water-ice erosive properties for the cold disk components in the previous section.

While the value of the {\it slope of the tensile strength curve} $s$ significantly affects the slope of the particle 
mass-distribution \citep{obrien03,gaspar12b}, it does not affect the decay of the fractional infrared emission 
to the level where we would observe offsets between the modeled and observed rates. However, we do have a best 
fit at its nominal value. 

{\it The effects of varying} $S$ {\it and} $Q_s$ are roughly the same as when varying $s$. As it turns
out, the exact value of the tensile strength law does not strongly influence the decay of the excess fractions in a population of debris disks. However, {\it choosing a higher value for $f_M$}, which gives the interpolation distance
between the erosive
and catastrophic collisional domains, does result in less acceptable fits. This is an arbitrarily chosen numerical
constant, and this analysis shows that choosing its value wisely is important. Based on these findings, we conclude
that for our cold disk models in the previous section, the changes to $\alpha$ and $b$ when assuming a water-ice
strength for the erosive collisions had a larger effect on the evolution than the changes to the catastrophic
collision properties of the tensile strength curve.

While {\it varying} $\eta$ (the initial particle mass distribution slope) of a single disk will have significant effects
on the timescale of its evolution (see Appendix), it does not strongly determine the timescale of the excess fraction
evolution of a population. To compensate for the offset in timescales, the average disk mass varies from
population to population (within an order of magnitude). Testing the actual value of the initial particle mass 
distribution is possible, by comparing the disk mass distributions predicted for each population to observations (such as in young clusters).
 
{\it Varying the maximum mass} $m_{\rm max}$ of the system did not have a large effect on the population synthesis fits above
$10^{18}~{\rm kg}$, which reinforces our previous statement that it is the dust density of the model that matters and not an absolute total mass or largest mass in the system, which are redundant variables. However, very low maximum mass systems 
($< 10^{18}~{\rm kg}$ -- $\approx 100~{\rm km}$ diameter) will result in decays that are inconsistent with our observations. 
This also has the important consequence, that the evolution of the planetary systems has to reach the point where bodies on
this size scale are common in order to have a ``successful'' collisional cascade.

{\it The radial distance of the model} ($R$) obviously is the dominant parameter. In section \ref{sec:24},
we showed that the best fit of our model to the observations is at $R\approx 4.5 - 5.5~{\rm AU}$, which
agrees with the thermal location predicted by \cite{morales11}. Here, we show the quality of the fits when
varying the radial distance between 2.5 and 10 AU. Placing the disks closer than 4 or further than 8 AU
yields a population decay that is inconsistent with the observations. This value can likely be modified 
to some extent by varying some of the other input variables of the model.

\section{Conclusions}
\label{sec:concl}

In this paper, we present a theoretical study of the evolution of debris disks, 
following their total disk mass ($M_{\rm tot}$), dust mass ($M_{\rm dust}$), and 
fractional 24 $\micron$ infrared emission ($f_{d(24)}$). We use the numerical code presented in Paper I that models the
cascade of particle fragmentation in collision dominated debris disk rings. 

Observational studies in the past decades have shown that the occurrence and strength of 
debris disk signatures fade with stellar age \citep[e.g.,][]{spangler01,rieke05,trilling08,carpenter09}. 
Analytic models of these decays explained them as a result of a steady-state (equilibrated) collisional cascade between 
the fragments \citep[e.g.,][]{spangler01,dominik03,wyatt07}, 
which results in a decay timescale proportional to $\propto t^{-1}$ for all model variables ($M_{\rm tot}$, $M_{\rm dust}$, $f_{d(24)}$).
Analysis of the observed decays of stellar populations, however, has shown that the dust mass and the fractional 
infrared emission -- the observable parameters -- decay less quickly \citep[e.g.,]{greaves03,liu04,moor11}.
Slower decays have also been modeled by complete numerical cascade models \citep[e.g.,][]{thebault03,lohne08,kenyon08}.
Numerical codes yield slower decays because they model the systems as relaxing in a quasi steady state, instead of 
in complete equilibrium. This means that mass is not entered at the high mass end 
into the system (like in an analytic model), but is rather conserved. 
The remaining discrepancies among the numerical models are results of the different collisional physics and processes 
modeled within them.\footnote{For a detailed description of the differences between the numerical models please
see Paper I.} 

Our calculations show that the evolution speed constantly varies over time and cannot
be described by a single value. Since the fractional infrared emission is a proxy for the dust mass, their
decays closely follow each other. At its fastest point in evolution, the total mass of our models decays as
$M_{\rm tot} \propto t^{-0.33}$, while the dust mass and fractional infrared emission of the single disk
decays $\propto t^{-0.8}$. At later stages in evolution these slow down to $\propto t^{-0.08}$ and 
$\propto t^{-0.6}$, respectively. These results are mostly in agreement with the models of \cite{kenyon08}.
We roughly agree with the dust mass decay predicted by the \cite{wyatt11} models up to the point where
Poynting-Robertson drag (PRD) becomes dominant in their models (although their models decay somewhat
faster than ours, possibly due to the constant effects of PRD).

We perform a population synthesis routine, assuming a log--normal probability distribution of initial disk 
masses. We calculate excess fraction decay curves, which we fit to the observed fraction of warm debris 
disks at a 10\% excess threshold at 24 $\micron$. Our fits show a good agreement between the calculated 
and observed decay rate of the fraction of debris disk sources around both solar and early-type stars, 
with initial mass ranges in agreement with the distribution of protoplanetary disk masses \citep{andrews05}.
We also analyze data from the MIPS/{\it Spitzer} and the DEBRIS and DUNES {\it Herschel Space Observatory} surveys.
Taking into account the non--uniform detection thresholds at these longer wavelengths, we also show
good agreement between the number of sources predicted to have an excess from our population
synthesis routines and that observed within these surveys. The best correspondence between models and 
observations requires grains that are relatively weak and have optical constants similar to those of water-ice composites.
However, a full range of grain properties was not explored.

There are a small number of bright debris disks at 24 $\micron$ around old stars that are not predicted 
by the simple decay from a log-normal starting distribution; they 
[HD 109085, HIP 7978 (HD 10647), HIP28103 (HD 40136), HIP40693 (HD 69830), $\eta$ Crv (HD 109085)] probably 
in part represent late-phase dynamical activity. Similarly, the model fails to fit the large excesses in the far 
infrared around old stars, again consistent with late-phase activity around a small number of stars. 

\acknowledgments

We thank K.\ Y.\ L.\ Su for substantial assistance in preparing the {\it Spitzer} data for this paper.
We thank Dr.\ Dimitrios Psaltis and Dr.\ Feryal \"Ozel for their contributions to the collisional cascade model
and the numerical code and also Dr.\ Michiel Min for providing the volatile mixture grain optical properties. 
Support for this work was provided by NASA through Contract Number 1255094 
issued by JPL/Caltech. Zolt\'an Balog is funded by the Deutsches Zentrum für Luf- und Raumfahrt (DLR). Partial 
support for this work was also provided for Zolt\'an Balog through Hungarian OTKA Grant \#K81966.

\appendix

\section{The system variables and their effects on the evolution of the collisional cascade}

As we have shown in Section 5, varying the parameters of the model can affect the results of the population
synthesis. Here, we analyze the effects of varying them on a single system. We summarize and describe
the variables of the model in Table \ref{tab:tabvar}.

\setcounter{table}{5}
\begin{deluxetable}{llr}
\tablecolumns{3}
\tabletypesize{\footnotesize}
\singlespace
\tablewidth{0pt}
\tablecaption{Numerical, Collisional, and System parameters of our model and their fiducial values\label{tab:tabvar}}
\tablehead{
\colhead{Variable} & \colhead{Description} & \colhead{Fiducial value} }
\startdata
\multicolumn{3}{c}{System variables}					 						\\
\hline\hline														
$\rho$		 & Bulk density of particles \dotfill									& 2.7 g cm$^{-3}$			\\
$m_{\rm min}$ 	 & Mass of the smallest particles in the system \dotfill						& 1.42$\times10^{-21}$ kg	\\
$m_{\rm max}$ & Mass of the largest particles in the system \dotfill							& 1.13$\times10^{22}$ kg		\\
$M_{\rm tot}$	 & Total mass within the debris ring \dotfill								& 1 $M_{\Earth}$ 			\\
$\eta_0$	 	 & Initial power-law distribution of particle masses \dotfill					& 1.87 					\\
$R$		 	 & Distance of the debris ring from the star \dotfill 						& 25 AU 					\\
$\Delta R$	 & Width of the debris ring \dotfill									& 2.5 AU 					\\
$h$		 	 & Height of the debris ring \dotfill								-			& 2.5 AU 					\\
Sp	 		 & Spectral-type of the star \dotfill								& A0 					\\
\hline															
\multicolumn{3}{c}{Collisional variables}										\\
\hline\hline														
$\gamma$	 & Redistribution power-law \dotfill									& 11/6 					\\
$\beta_X$	 	 & Power exponent in X particle equation \dotfill						& 1.24 					\\
$\alpha$		 & Scaling constant in $M_{\rm cr}$ \dotfill							& $2.7\times10^{-6}$ 		\\
$b$			 & Power-law exponent in $M_{\rm cr}$ equation \dotfill						& 1.23 					\\
$f_M$		 & Interpolation boundary for erosive collisions \dotfill						& $10^{-4}$ 				\\
$f_Y$		 & Fraction of $Y/M_{\rm cr}$ \dotfill									& 0.2 					\\
$f_X^{\rm max}$  & Largest fraction of $Y/X$ at super catastrophic collision boundary \dotfill 			& 
0.5 					\\
$Q_{\rm sc}$	 & Total scaling of the $Q^{\ast}$ strength curve \dotfill						& 1 						\\
$S$			 & Scaling of the strength regime of the $Q^{\ast}$ strength curve \dotfill			& $3.5\times10^{7}$ erg/g 	\\
$G$			 & Scaling of the gravity regime of the $Q^{\ast}$ strength curve \dotfill			& 0.3 erg cm$^3$/g$^2$		\\
$s$			 & Power exponent of the strength regime of the $Q^{\ast}$ strength curve \dotfill		& -0.38 					\\
$g$			 & Power exponent of the gravity regime of the $Q^{\ast}$ strength curve \dotfill 		& 1.36  					\\
\hline															
\multicolumn{3}{c}{Numerical parameters}										\\
\hline\hline														
$\delta$	 	& Neighboring grid point mass ratio \dotfill							-		& 1.104			 		\\
$\Theta$	 	& Constant in smoothing weight for large-mass collisional probability\dotfill			& $10^6 m_{\rm max}$		\\
$P$		 	& Exponent in smoothing weight for large-mass collisional probability\dotfill			& 16				
\enddata
\end{deluxetable}

\subsection{Evolution of the system mass}

\begin{figure*}
\begin{center}
\includegraphics[angle=0,scale=0.8]{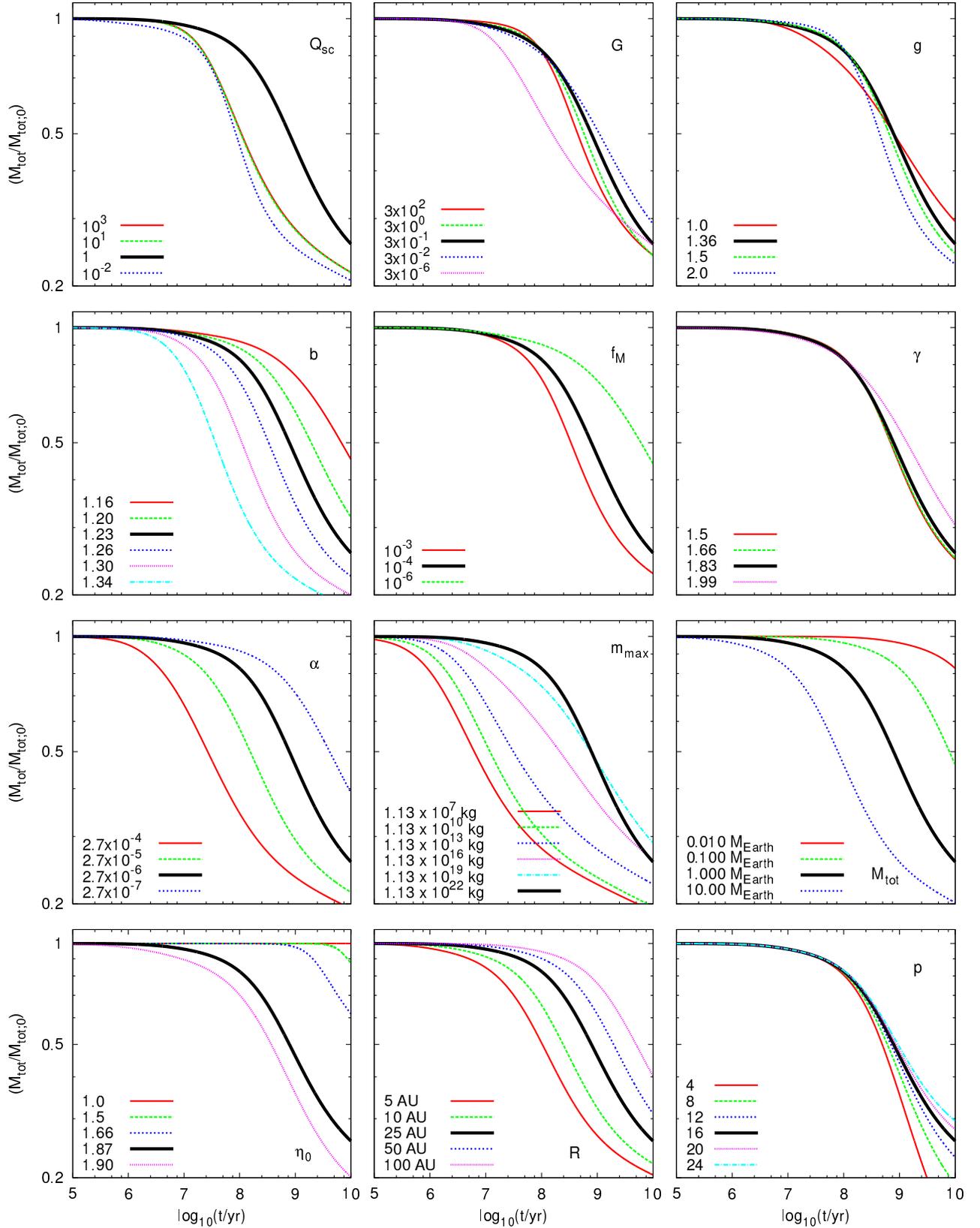}
\caption{Evolution of the total disk mass as a function of selected parameters that have the largest effect on the timescale
of the evolution. The numerical variable $p$ modifies the smoothing function of the collisional cross section of the largest 
bodies in the system. The smoothing function only varies the evolution of the total mass (shown here), but does not affect 
the evolution of the dust mass or the fractional infrared emission.}
\label{fig:mtot}
\end{center}
\end{figure*}

\begin{figure*}
\begin{center}
\includegraphics[angle=0,scale=0.8]{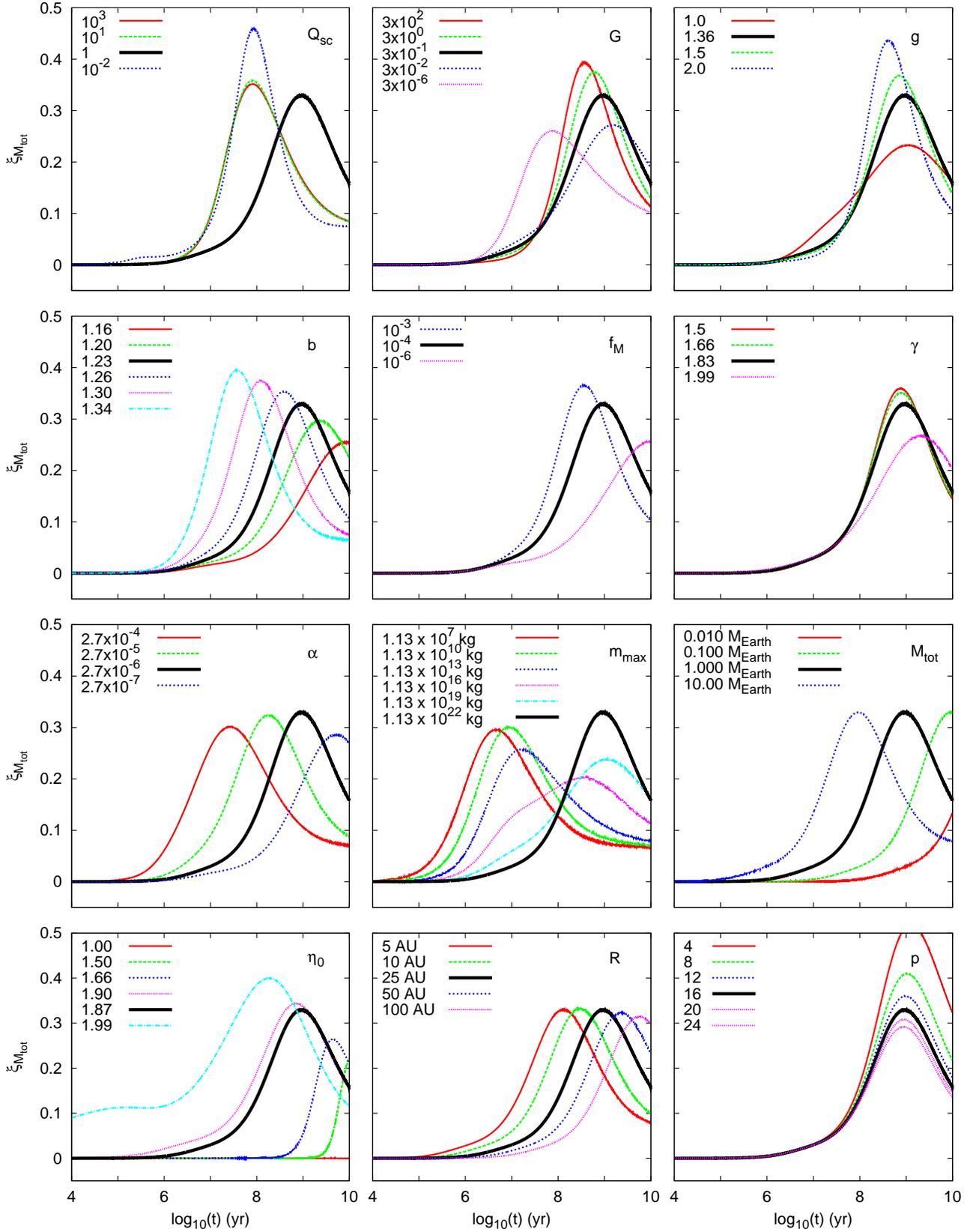}
\caption{The evolution of the power exponent of time in the decay of the total system mass [$M_{\rm tot}(t) \propto t^{-\xi}$]
as a function model variables for decay curves shown in Figure \ref{fig:mtot}. At its fastest point, our reference model
decays with $\xi=0.33$, while all models reach a fastest point between $0.3 < \xi < 0.4$.
The evolution of the power exponent is characteristic for all models, with an acceleration in evolution up to a certain point, from whereon the evolution
of the total mass slows down.}
\label{fig:mtot_slope}
\end{center}
\end{figure*}

We show the total mass decay curves as a function of model variables in Figure \ref{fig:mtot} and the evolution 
of the power exponent of time in the decay of the total mass [$M_{\rm tot}(t) \propto t^{-\xi}$] as a function of these 
collisional variables in Figure \ref{fig:mtot_slope}. The figures include plots for the twelve variables that 
have the largest effect on the evolution, out of the total twenty-four variables (see Paper I).
These decays are compared to that of our reference model, detailed in section \ref{sec:mtot}.

In our code, we use the models of \cite{benz99} to estimate the collision tensile strengths of particles, written as
\begin{equation}
Q^{\ast}(a) = 10^{-4} \frac{\rm J~g}{\rm erg~kg} Q_{\rm sc}\left[S \left(\frac{a}{1~{\rm cm}}\right)^{s} + G \rho \left(\frac{a}{1~{\rm cm}}\right)^{g}\right]\;,
\end{equation}
where $a$ is the target particle's radius, $Q_{\rm sc}$ is the total scaling of the curve, $S$ is the scaling of the 
curve in the ``strength dominated'' regime, $s$ is the power exponent of the target radii in the ``strength dominated''
regime, $G$ is the scaling of the curve in the ``gravity dominated'' regime, $\rho$ is the bulk density of the 
particles, and $g$ is the power exponent of the target radii in the ``gravity dominated'' regime. Of these, we 
show the effects of varying $Q_{\rm sc}$, $G$, and $g$, as varying $S$ and $s$ will not have a significant 
effect on the decay of the total mass, 
because they influence the low mass end of the distribution. 
Increasing or decreasing the total scaling will speed up the evolution of the total mass. 
Decreasing the total scaling of the tensile strength 
curve will soften the materials, resulting in a faster decay. Increasing it, however, will strengthen
the materials, which will make the largest bodies ``indestructible'', resulting in a faster decay in the number
of bodies just below the high mass end. A similar effect can be seen when $G$ is varied.

The total mass cratered in an erosive collision is calculated in our model by applying the experimental results of 
\cite{koschny01a,koschny01b}. This mass is a function of $\alpha$ (scaling constant) and the projectile's 
energy to the $b$ power. 
Variations in these constants will affect how quickly the largest bodies erode and subsequently, the 
evolution of the total disk mass. When softer material properties are used ($\alpha$ and $b$ increase), 
the decay is quicker, for example meaning debris disks composed of ice are likely to disappear in a shorter timescale 
than rocky debris disks.

The \cite{koschny01b} formula for cratered mass in an erosive collision is only valid for relatively small 
cratered masses. The cratered mass given by the formula can exceed $M/2$ even below the Erosive/Catastrophic 
collision boundary. We thus interpolate the cratered mass from $f_M = M_{\rm cr}/M$ to the boundary via methods 
given in Paper I. Assigning it a very small value basically eliminates the 
erosive formula of \cite{koschny01b} and uses an interpolative formula for the entire domain. However, a larger value 
is likely to overestimate the cratered mass in an erosive collision near the erosive/catastrophic collision boundary. Our approach was to use a conservative value within these extremes.

The number densities of fragments created in collisions in our model follow a power-law distribution. The slope of this 
distribution is given by $\gamma$, and only very minor effects can be seen when varying its value. 
The actual redistribution function has been a long researched topic within collisional systems, with 
some research showing that double or even triple power-law functions 
are the best to describe the fragment distributions \citep{davis90}. According to our models, as long as the 
distribution function is within reasonable limits ($\gamma < 1.99$ - mass is concentrated in the largest fragments), 
there is not much difference in the decay of the total mass when varying its value.

The total mass within the disk, $M_{\rm tot}$, sets the scaling of the particle size distribution 
(as do $m_{\rm max}$ and the volume of the disk). When scaling the initial total mass in the system, with
all other parameters fixed, the evolution of the total mass is shifted in time, with the systems
reaching their points of fastest decay at later points in time. This property is used in our population
synthesis calculations in section \ref{sec:fit}.

The decay is dependent on the mass of the largest body $m_{\rm max}$, which is 
usually arbitrarily chosen in the numerical models. This shows in our calculations in Section 
\ref{sec:constr}, where going below a largest body mass of $10^{18}~{\rm kg}$ ($\approx 100~{\rm km}$ diameter) 
will result in decays that are inconsistent with our observations. When testing this for a single system, 
we set the total mass of the system to a value that yielded the same scaling of particle densities as the fiducial 
model had. This way we guaranteed that our calculations were only testing how varying the cutoff of the 
mass distribution affects the evolution. 

The slope of the initial distribution, $\eta$, determines the number of dust particles when the
collisional cascade is initiated. Our convergence tests (Paper I) have shown that the 
systems will reach collisional equilibrium from all initial distribution slope values. However,
the time when the system reaches equilibrium will depend on the value of $\eta$. A system will be 
able to reach equilibrium from slope values lower than the steady-state cascade distribution faster
than from steeper slopes, as it is easier to produce and build up dust sizes, than to remove
the large massive particles from the highest masses.

One of the most important system variables is $R$, the distance of the disk from the central
star. This parameter has many effects, as it sets the collisional velocity, thus the collisional
energy of the particles and their collisional rate. It will also set the removal
timescale for the blowout particles and is a variable in the volume of the disk, thus it sets the
number density of the particles in the disk as well. Increasing the the radial distance will
decrease the evolution rate of the disks, as shown in our Figures \ref{fig:mtot} and \ref{fig:mtot_slope},
with the fastest evolution setting in at later points in time.

The last parameter we analyze is $p$, which is a variable that sets the smoothing function
of the collisional rates for the largest bodies in the system (Paper I). Its value only affects
the evolution of the largest masses, thus also the evolution of the total mass in the system. 

\subsection{Evolution of the dust mass and fractional infrared emission}

As we have shown before, the fractional infrared emission is a proxy of the dust 
mass in the system, meaning the decay curves and the analysis we give for the fractional infrared 
emission are generally identical to the one we would give for the dust mass in the system. For 
said reasons, we omit the plots for the evolution of the dust mass.

The emission of the particles depends on their temperatures, their sizes, and material and 
wavelength dependent optical properties, such as their absorption coefficients. We assume 
a \cite{castelli03} intensity emission model for the stars and astronomical silicate optical 
constants for the particles \citep{draine84}, when calculating their equilibrium temperatures 
and emission.

\begin{figure*}
\begin{center}
\includegraphics[angle=0,scale=0.8]{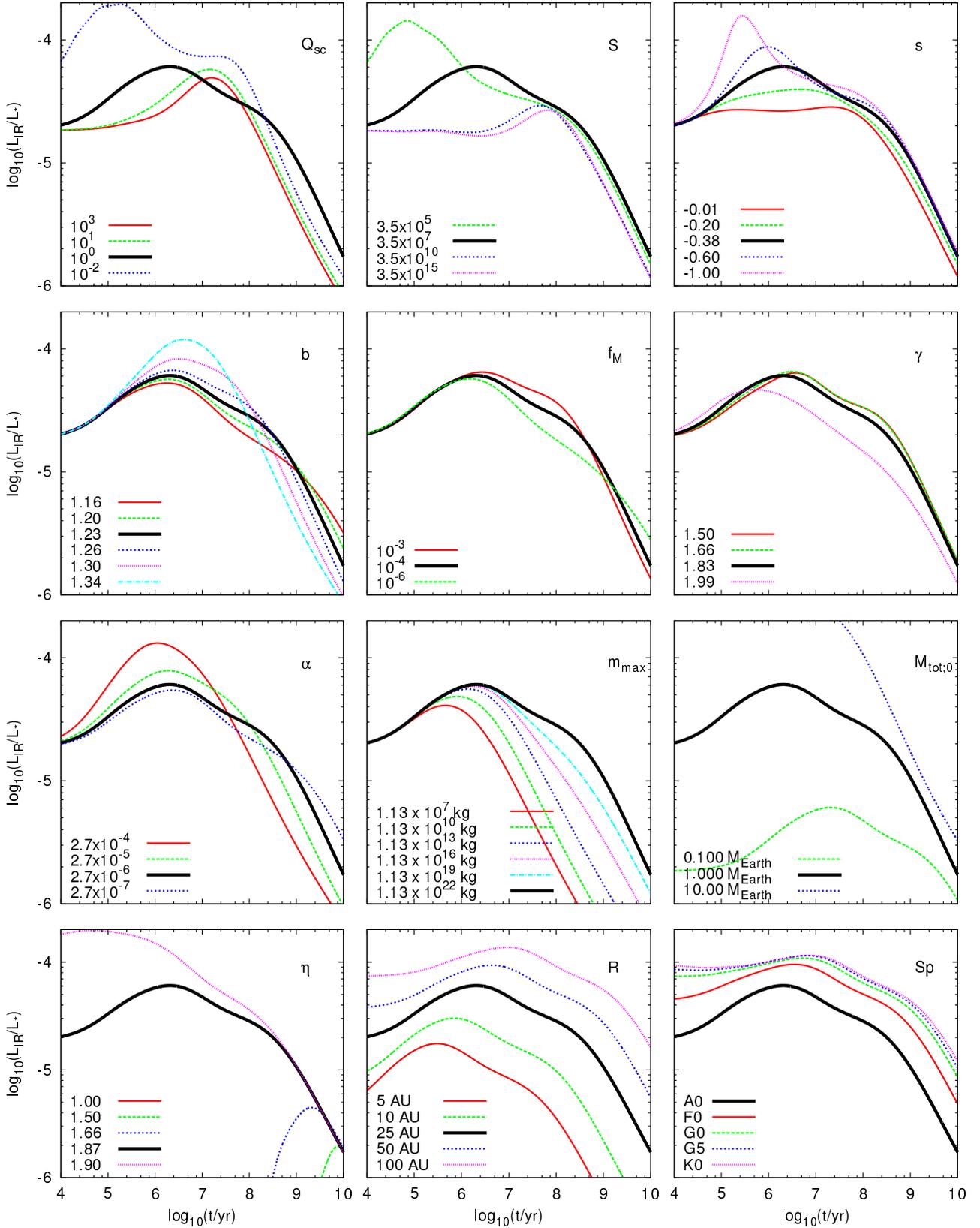}
\caption{Evolution of the fractional infrared luminosity 
as a function of selected parameters that have the largest effect on the timescale of its evolution. 
The characteristic bump seen in the evolution of the dust mass is reflected in the evolution of the infrared 
emission as well. The bump is followed by a drop in emission, which follows the same power-law as the drop in 
dust mass. Systems generally reach the quasi steady state decay at $\sim 100~{\rm Myr}$, although variations in 
this are seen as a function of model parameters.}
\label{fig:lir}
\end{center}
\end{figure*}

\begin{figure*}
\begin{center}
\includegraphics[angle=0,scale=0.8]{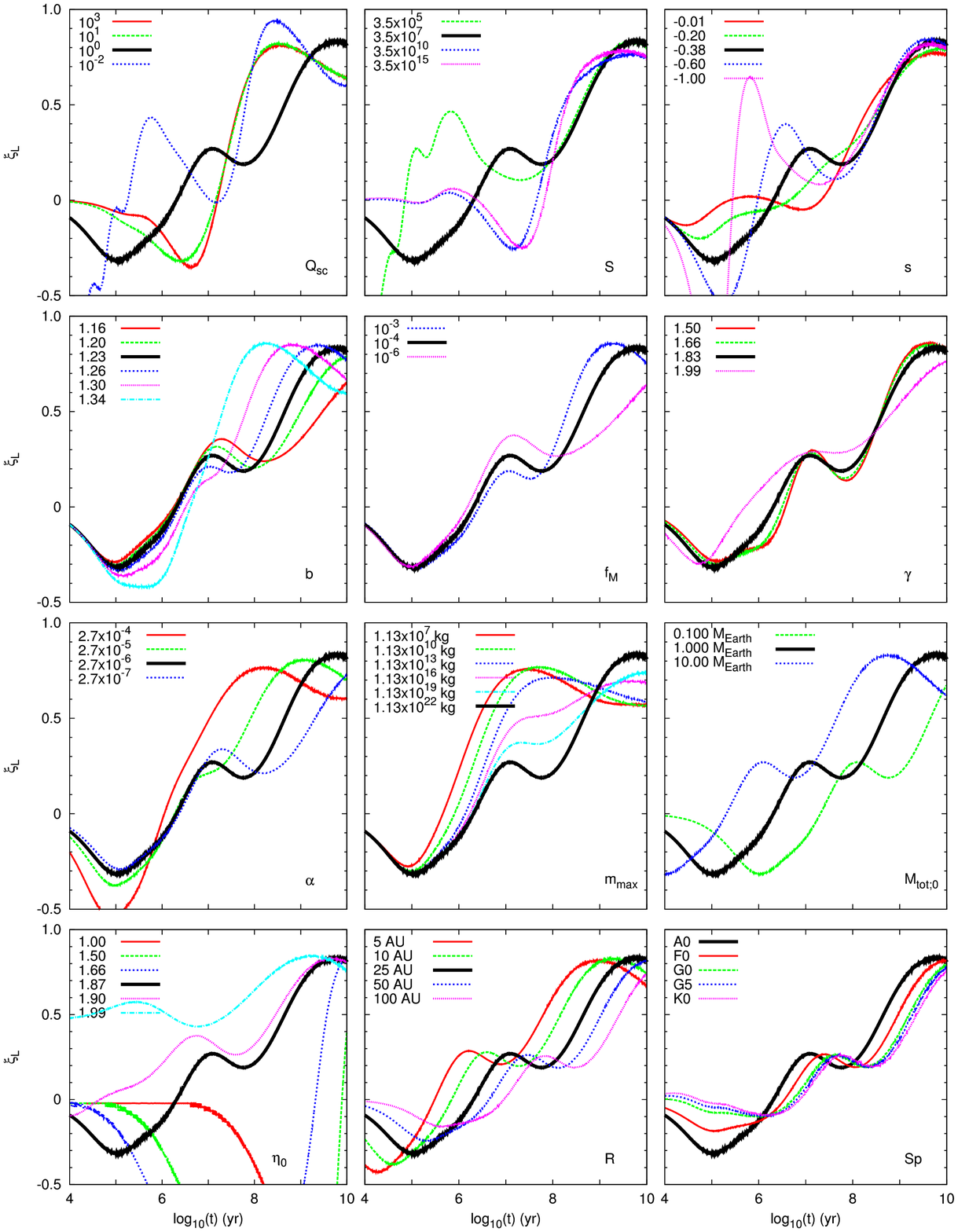}
\caption{The evolution of the power exponent of time in the decay of the fractional infrared luminosity
as a function model variables for decay curves shown in Figure \ref{fig:lir}. At its fastest point, the infrared emission our reference model
decays with $\xi=0.8$, while all models reach a fastest point between $0.6 < \xi < 0.9$.
These model results generally agree with observations of disk decay.}
\label{fig:lir_slope}
\end{center}
\end{figure*}

We analyze the same parameters as in the previous subsection, with the exception of $G$, $g$, 
and $p$, which are replaced by $S$, $s$, and $\delta$. In Figure \ref{fig:lir}, we show the 
decay of the fractional infrared emission as a function of the model variables that have the largest 
effect on it, while in Figure \ref{fig:lir_slope}, we show the power exponent of time in the decay.
These figures can be compared the the evolution of the infrared emissions of our reference model,
which is plotted with a thick solid line in the Figures and also analyzed in section \ref{sec:lir}.

The variables of the tensile strength curve that determine the strengths of the gravity dominated 
larger bodies ($G$, $g$) do not affect the evolution of the dust distribution, while the variables 
that determine the tensile strengths of the smaller particles ($S$, $s$) do. Increasing or decreasing the scaling
of the tensile strength law ($Q_{\rm sc}$) increases the evolution speed for the dust mass, and thus
the fractional infrared emission. At increased material strengths the quick decay of the largest bodies 
affects the evolution of the dust mass, while for softer materials a general faster decay of the entire distribution 
can be seen (see Figure 4 in Paper II). However, only significant decreases in the strength scaling $S$
will have noticeable effects in the evolution of the fractional infrared emission. Increasing the steepness of the tensile
strength law $s$ will shift the evolution in time. Of all collisional variables, arguably $b$ and $\alpha$ 
are the most important. As expected, using softer erosive material properties (larger $b$ and $\alpha$)
speeds up the evolution of the dust mass (and with that the evolution of the fractional infrared emission).

Changes in $f_M$ and $\gamma$ affect the evolution of the fractional infrared emission
similarly to that of the total mass. 
Increasing the largest body in the system ($m_{\rm max}$) slows down 
the evolution of the collisional cascade, with models reaching their peak dust mass evolution 
at later stages, while increasing the total mass $M_{\rm tot}$ in the system 
will speed the evolution of the system, with higher total mass systems reaching 
their peak evolutionary point earlier on. 
Systems initiating their collisional 
cascades with varying initial mass-distribution slopes ($\eta_0$) will reach their 
quasi steady state dust mass decay (the peak of evolution speed) roughly
at the same time, even though the beginning of the evolution is dependent on 
the slope. Debris rings located at different radial distances ($R$) will evolve with speeds
associated with their orbital velocities, shifting the onset of their quasi steady state decay
to later points in time for disks at larger radial distances. 

Since $p$ is the smoothing function of the largest bodies, it also does not affect 
the evolution of the dust mass; however, the neighboring grid point mass ratio ($\delta$) 
will be numerically important. In Figure \ref{fig:lir},
we show that our models converge in dust mass decay at around 400-800 grid points,
while using a less dense grid will result in numerical errors.

\subsection{Conclusion}

Our analysis above has revealed that erosive collisions are dominant in shaping the evolution 
of a debris disk. The evolution speed of our model is determined primarily by the variables ($\alpha$ and $b$)
of the cratered mass equation, when considering fixed system variables. This is not that
surprising, considering that $b$ also was found to be dominant in determining the mass-distribution 
slope (Paper II), and that our population synthesis analysis in section \ref{sec:constr} also revealed 
that our fits are sensitive to the values of $\alpha$ and $b$. The evolution is
much less dependent on the catastrophic tensile strength than on the erosive, which is surprising,
considering the dependence of the particle mass-distribution slope on $s$ \citep[][Paper II]{obrien03}.

The measurements of \cite{koschny01a,koschny01b} give the value of $\alpha$ for silicates as 
$2.7\times10^{-6}~{\rm kg~J}^{-1}$, and $6.2\times10^{-5}~{\rm kg~J}^{-1}$ for ice; and
a value of 1.23 for $b$. Measurements by \cite{hiraoka08} yield a $b$ value of 1.15, which is in
agreement with the value given by \cite{koschny01a,koschny01b} and yields an even better fit
for our population synthesis constraint in section \ref{sec:constr} ($\alpha$ values cannot 
be compared as the papers used slightly different equations).

Of the system variables, the evolution will most strongly depend on $\eta_0$, $R$, and $m_{\rm max}$.
The evolution converges above $m_{\rm max} = 1\times10^{19}~{\rm kg}$ ($\approx 200~{\rm km}$ diameter),
which most systems likely achieve (considering the asteroid sizes in our Main Asteroid Belt), making
this variable less important for realistic conditions. Although $\eta_0$ is difficult to constrain,
it is likely that the system will form with a mass-distribution slope with a value close to its
quasi steady state solution. However, even if a system does not, its evolution still can adequately reproduce 
our observations according to our population synthesis calculations in section \ref{sec:constr}.

The radial distance of the disk is the overall dominant parameter in determining the evolution
of a single disk, when all realistic conditions are considered. It influences the evolution by
three independent effects, all acting in the same direction. At larger radii, the collisional
velocity will be lower (thus the collisional energy will be lower), which lowers the effective mass range a 
particle can interact with. The reduced collisional velocity also reduces the collisional rate.
Finally, an increase in radial distance increases the effective volume the disk encompasses (for
the same amount of mass and disk aspect ratio), also resulting in reduced collisional rates.

\setcounter{table}{1}
\clearpage
\LongTables
\begin{landscape}
\tabletypesize{\scriptsize}
\begin{deluxetable*}{llr|rrr|rrrrrr|rrr|rcc}
\tablecolumns{18}
\tablewidth{0pt}
\tablecaption{Photometry of the DEBRIS and DUNES surveys.\label{tab:midIR}}
\tablehead{
\colhead{} & \colhead{} & \colhead{} & \colhead{} & \colhead{} & \colhead{} & \multicolumn{6}{c}{MIPS} & \multicolumn{3}{c}{PACS} & \colhead{Far} & \colhead{} & \colhead{} \\
\colhead{Name} & \colhead{Sp.} & \colhead{Age} & \colhead{P$_{24}$} & \colhead{P$_{70}$} & \colhead{P$_{100}$} & \colhead{$F_{\rm 24}$} & \colhead{$\sigma_{\rm 24}^{\ast\ast}$} & \colhead{$R_{\rm 24}^{\Box}$}  
& \colhead{$F_{\rm 70}$} & \colhead{$\sigma_{\rm 70}$} & \colhead{$\chi_{\rm 70}$} & \colhead{$F_{\rm 100}$} & \colhead{$\sigma_{\rm 100}^{\ddagger}$} & \colhead{$\chi_{\rm 100}$} 
& \colhead{IR} & \colhead{Age$^{\Diamond}$} & \colhead{Age ref.}\\
\colhead{} & \colhead{type} & \colhead{(Gyr)} & \colhead{(mJy)} & \colhead{(mJy)} & \colhead{(mJy)} & \colhead{(mJy)} & \colhead{(mJy)} & \colhead{(mJy)} & \colhead{(mJy)} & \colhead{(mJy)} & \colhead{} 
& \colhead{(mJy)} & \colhead{(mJy)} & \colhead{} & \colhead{Exc?} & \colhead{Flag\phm{$^{\Diamond}$}} & \colhead{}}
\startdata
\multicolumn{18}{c}{DEBRIS survey (HD sources only)}\\
\hline\hline
HD000038		&	K6V	&	1.15	&	60	&	6.63	&	3.33	&	-	&	-	&	1.00	&	-	&	-	&	-	&	11.14\phm{$^{\checkmark\bullet}$}  	   &	   4.96    &	   1.56    &	   N	   &	   2	   &	   1;2\\   
HD000739		&	F5V	&	2.15	&	161	&	18.01	&	8.71	&	163	&	1.63	&	1.01	&	20.11	&	2.59	&	0.76	&	21.19\phm{$^{\checkmark\bullet}$}  	   &	   7.87    &	   1.57    &	   N	   &	   1	   &	   3\\  																																								   
HD001237		&	G8V	&	0.30	&	84	&	9.28	&	4.59	&	84	&	0.85	&	1.01	&	11.81	&	2.00	&	1.21	&	-4.26\phm{$^{\checkmark\bullet}$}  	   &	   3.38    &	   -2.61   &	   N	   &	   3	   &	   2;3;4;5;6\\  																																								   
HD001326		&	-	&	-	&	236	&	26.08	&	12.42	&	-	&	-	&	-	&	-	&	-	&	-	&	17.45\phm{$^{\checkmark\bullet}$}  	   &	   5.11    &	   0.97    &	   N	   &	   -	   &	   -\\  																																								   
HD001404		&	A2V	&	0.45	&	129	&	14.27	&	7.06	&	155	&	1.56	&	1.20	&	43.82	&	6.12	&	4.55	&	27.78\phm{$^{\checkmark\bullet}$}  	   &	   2.68    &	   6.86    &	   Y	   &	   2	   &	   7\\  																																								   
HD001581		&	F9.5V	&	3.82	&	557	&	61.55	&	29.23	&	557	&	5.57	&	1.00	&	71.20	&	5.73	&	1.43	&	45.25\phm{$^{\checkmark\bullet}$}  	   &	   4.63    &	   3.11    &	   N	   &	   3	   &	   3;5;6;8\\																																									   
HD001835		&	G3V	&	0.44	&	80	&	8.84	&	4.37	&	-	&	-	&	1.06	&	-	&	-	&		&	22.41\phm{$^{\checkmark\bullet}$}  	   &	   6.42    &	   2.77    &	   N	   &	   3	   &	   2;3;5;6;9;10;11;12;13;14\\
\enddata
\end{deluxetable*}
\clearpage
\end{landscape}

\end{document}